\documentclass[a4paper,10pt,twocolumn,preprint,3p, authoryear]{elsarticle}

\journal{Journal of \LaTeX\ Templates}

\usepackage{lineno}
\usepackage{hyperref}

\modulolinenumbers[5]
\usepackage{color}

\usepackage[T1]{fontenc} 

\usepackage{epstopdf}
\usepackage{subcaption}
\usepackage[toc,page]{appendix}
\usepackage[T1]{fontenc}
\usepackage{inputenc}
\usepackage{amsmath}
\usepackage{float}
\usepackage{dblfloatfix}
\usepackage{textcomp}

\begin{document}
\journal{TexExchange}

\begin{frontmatter}

\title{\boldmath Prospects of Probing the Radio Emission of Lunar UHECRv Events}

\author[mysecondaryaddress]{A. Aminaei\corref{mycorrespondingauthor}}
\cortext[mycorrespondingauthor]{Corresponding author}
\ead{amin.aminaei@physics.ox.ac.uk}

\address[mysecondaryaddress]{Department of Physics, University of Oxford, OX1 3RH, UK}

\author [b]{L. Chen}
\address[b]{National Astronomical Observatories, Chinese Academy of Sciences, Beijing, China}

\author [c1,c2]{H. Pourshaghaghi}
\address [c1]{Department of Astrophysics/IMAPP, Radboud University, Nijmegen, P.O. Box 9010, 6500 GL Nijmegen, The Netherlands}
\address [c2]{Eindhoven University of Technology, PO Box 513, 5600 MB Eindhoven, The Netherlands}

\author [e]{S. Buitink}
\address [e]{Astrophysical Institute, Vrije Universiteit Brussel, Pleinlaan 2, 1050 Brussels, Belgium}

\author [c1]{M. Klein-Wolt}

\author [g]{L. V. E. Koopmans}
\address [g]{Kapteyn Astronomical Institute, University of Groningen, PO Box 800, NL-9700 AV Groningen, The Netherlands}

\author [c1,h2]{H. Falcke}

\address [h2]{ASTRON, Oude Hoogeveensedijk 4, 7991 PD Dwingeloo, The Netherlands}

\begin{abstract}

Radio detection of Ultra High Energetic Cosmic Rays and Neutrinos (UHECRv) which hit the Moon has been investigated in recent years. In preparation for near-future lunar science missions,  we discuss technical requirements for radio experiments onboard lunar orbiters or on a lunar lander. We also develop an analysis of UHECRv aperture by including UHECv events occurring in the sub-layers of lunar regolith.  It is verified that even using a single antenna onboard lunar orbiters or a few meters above the Moon's surface, dozens of lunar UHECRv events are detectable for one-year of observation at energy levels of $10^{18}$eV to $10^{23}$eV. Furthermore, it is shown that an antenna 3 meters above the Moon's surface could detect lower energy lunar UHECR events at the level of $10^{15}$eV to $10^{18}$eV which might not be detectable from lunar orbiters or ground-based observations.  
   
\end{abstract}

\begin{keyword}
\texttt{ UHECRv, Cosmic Rays, Cosmic Neutrinos, Lunar Lander, Lunar Orbiter, Space Radio Experiment}

\end{keyword}

\end{frontmatter}
\linenumbers
\section{Introduction}
\label{sec:intro}

\nolinenumbers
Radio emission from the cascades of energetic particles has been studied intensively in recent years. Particle accelerators such as Stanford Linear Accelerator Center (SLAC)  produces radio emission from particle cascades \citep{saltzberg}, \citep{slac}.  Along with lab experiments, radio experiments for studying the cosmic energetic particles in air showers and on the lunar surface are being developed. These experiments help to investigate the fundamental questions about the UHECRv such as their origin and the acceleration mechanism. The Moon has been long known as a detector for UHECRv events. The unique properties of Moon regolith such as very low conductivity and low attenuation makes it an ideal environment for detection of coherent radiation based on the Askaryan Effect \citep{ref1}. We refer to this radio emission, also known as Cherenkov-like radiation, as Askaryan radiation through the paper(as stated in \citep{askaref1} and \citep{askaref2}). 
Askaryan radiation is spread over a broad spectrum covering microwave frequencies (GHz band corresponds to cm wavelengths) in dielectric solids, but it may also reach a peak at lower frequencies within tens of MHz \citep{Scholten}. Most of radio UHECRv experiments at MHz regime operate, however, at frequencies higher than 100 MHz where dispersion in the Earth's ionosphere, and the Galactic background noise, become low. Also the dimension of antennas becomes reasonably smaller at higher frequencies. In our analysis frequencies of 1.5 GHz and 150 MHz represent the GHz and MHz frequency regimes. From theoretical estimates \citep{zas} and the SLAC experiments \citep{saltzberg} can be understood that the dominant mechanism for producing lunar UHECRv radiation is charge excess. This is mainly due to the absence of strong magnetic fields, such as the Earth's magnetic field, which is a key element in radio emission of air showers.  The impact of energetic charged particles with lunar regolith
generates electromagnetic pulses which develop and propagate as a cascade of electric currents through the layers of
lunar regolith and lunar exosphere. Antennas onboard lunar orbiters, a lunar lander or a ground-based
array can detect these events by measuring corresponding electric
fields (also known as the lunar Askaryan technique \citep{DZ}). In this paper we generalize the analytical methods in the literature for calculating the UHECRv apertures of ground based arrays so it can be used also for lunar orbiter experiments as well as for antennas on the Moon's surface. For the latter we modified the method and included the events occurring in the sub-layers of lunar regolith. The results are used to estimate the number of events that can be detected for various radio experiments for a one-year observation.

\begin{itemize}
\item Radio Experiments of Lunar UHECRv Emission

\item Analysis of Lunar UHECRv Events
\item Categorization of Lunar UHECRv Events
\item Technical Requirements of Future Lunar Radio Experiments
\item Summary and Conclusion
\end{itemize}
\begin{figure}

\includegraphics[width=0.45 \textwidth]{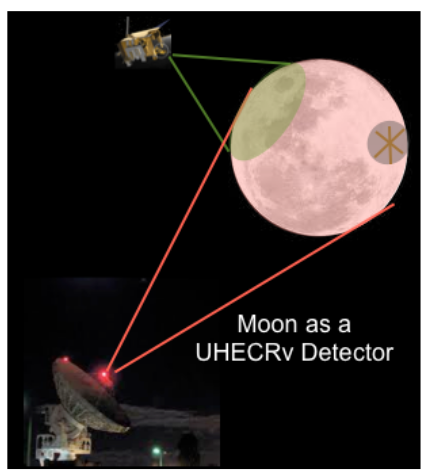} 

\caption{Observations of lunar UHECRv events using ground-based, lunar orbiter and Moon's surface radio experiments. Dimensions are not to scale.}
\label{fig:i}
\end{figure}
\section{Radio Experiments of Lunar UHECRv Emission} 
Radio detection of lunar UHECRv emission is limited by antenna sensitivity  which depends on the effective collecting area of antennas. Thus large ground-based antenna arrays provide the highest sensitivity, however, the measured electric field of lunar UHECRv emission is weak because of the distance. Another limiting factor is the physical area of the Moon which is illuminated by antennas. As shown in Fig.\ref{fig:i}, the physical area depends on the beamwidth of the antenna and its distance from the lunar surface.  The larger the area the higher chance of detection of lunar UHECRv events. A probability function then relates the physical area to the actual aperture for detection of UHECRv events. The probability function depends on the electromagnetic properties of UHECRv. It also depends on system parameters that we will shortly present.
There are various experiments capable of detection of Lunar UHECRv events \citep{bray}. We compare the expected outcome of the LOw Frequency ARray (LOFAR) \citep{lofar} and the Square Kilometer Array (SKA) \citep{ska}  telescopes with observations from a lunar orbiter or the lunar surface (e.g. antenna onboard a lunar lander).
\subsection{Lunar UHECRv Experiments}
Orbital Low Frequency ARray (OLFAR ) \citep{olfar} and Lunar Orbiter Radio Detector (LORD) \citep{Gusev}, \citep{lordd} are examples of planned Lunar Orbiter Observations of UHECRv events. We analyze the UHECRv detections from different altitudes and compare the expected results with results of simulations of UHECv events from a tripole antenna at 100 MHz onboard lunar orbiter satellites \citep{Stal}.\\ %
For observations of UHECRv events on the lunar surface, the analysis is done for an individual antenna onboard a lunar lander. It is based on the preliminary analysis \citep{icrc} of the Lunar Radio eXplorer (LRX) experiment. LRX is a dedicated radio experiment initially designed for the proposed European lunar lander mission. LRX includes a tripole antenna and a sensitive digital receiver in the frequency range of 5kHz-100MHz. The experiment was designed to observe radio emissions on the lunar surface including UHECRv radiation. Technical requirements and science cases of LRX are described in \citep{KleinW, Zarka}. \\ An individual antenna is used for both the lunar orbiter and the lunar surface experiments in this study. The analysis can be extended for multiple lunar orbiter antennas or for an array of antennas on the Moon's surface. \\
For all observations, UHECRv events can be detected if the maximum electric field of Askaryan radiation is equal to or greater than the minimum electric field detectable by the receiver. For Askaryan radiation, we use the formula presented in \citep{fdtd3} and \citep{ref01}. Furthermore, we extend the analysis by applying formulas in \citep{fdtd3} for both the lunar regolith and sub-regolith. Based on these formulas,  the electric field caused by Askaryan radiation depends on the energy of particles, the distance from the observer and frequency of observation. The minimum detectable electric field is defined \citep{glue} by characteristics of the antenna and receiver, and electromagnetic properties of the dielectric (e.g. lunar regolith and sub-regolith). 
\begin{eqnarray}
\label{eq:1}
\hspace*{-0.7cm}{\epsilon_{\rm min}} = N_{\sigma}. (2k_{b})^{1/2}. (\frac { T_{\rm sys} }{A_{e}\ \Delta v}) ^{1/2} \, . \, (\frac {Z_{0}}{n_{r}}) ^{1/2}  
\end{eqnarray}

The dependence of the minimum detectable electric field ($ \epsilon_{\rm min},$V/m)  on system parameters and lunar environment is shown in Eq.\ref{eq:1}.  
$N_{\sigma}$ is the minimum number of standard deviations needed to reject statistical noise pulses and is set to 5 in our analysis and $k_{b}$ is Boltzmann's constant. $T_{\rm sys}$ is system noise temperature, $A_{e}$ is the effective collecting area of the antenna(s) and $\Delta v$ is the bandwidth of the receiver. In the last term ${Z_{0}}$ and ${n_{r}} $ represent characteristic impedance ($Z_{0}=377 \; \Omega$) and the refractive index of the lunar surface respectively.
The radiation transmission occurs when the ray reaches the surface at an angle of incidence which is bigger than the complement of the Cherenkov angle ($\theta_{c}= \cos^{-1}(1/n_{r}$)). For plane waves, the transmission rate is defined with the transmission coefficient ($t_{||}$) as follows: \citep{Williams}


\begin{equation}
\label{eq:10}
{  t_{||}=(n_{r} . \cos \beta / \cos \beta_{0}. (1-r_{||}^{2}))^{0.5} \quad \quad \quad \quad } \\
\end{equation}



Where $\beta$ is the angle of incidence to the Moon's surface normal and $\beta_{0}$ is the angle of refraction relative to the
normal (outside the Moon's surface) as the rays pass through the Moon's surface into free space or through the sub-regolith into the lunar regolith. $r_{||}$ is the field reflection coefficient and the polarisation is assumed to be in the plane of incidence. A further numerical technique 
has been developed in \citep{Williams} to convert this plane-wave transmission coefficient to the coefficient which meets the criteria for UHECRv rays within the observed solid angle. 

A constant approximation of $t_{||}=0.6$ has been used in \citep{ref01} which we also used for our analysis.  

\subsection{Frequency of UHECRv Observation}
UHECRv radio emission covers a broad frequency range from kHz to GHz regime. Ground based radio experiments of UHECRv are common in MHz bands.  LOFAR, SKA Low and Westerbork Synthesis Radio Telescope (WSRT lower band, NuMoon \citep{numoon}) are examples of MHz radio telescopes for detection of lunar UHECRv events.  SKA Mid2, the higher band of the WSRT array, the Goldstone Lunar Ultra-high-energy Neutrino Experiment (GLUE, \citep{glue}), the Lunar Ultra-high-energy Neutrino Astrophysics with
the Square Kilometre Array (LUNASKA, \citep{lunaska}), the Radio EVLA Search for Ultra-high-energy Neutrinos (RESUN, \citep{cone}) and Five-Hundred-Meter Aperture
Spherical Radio Telescope (FAST) \citep{fast} are examples of GHz experiments for UHECRv events. A series of past and near-future lunar radio experiments for detection of UHECRv events are described in \citep{bray}. \\
In this study, we analyse the UHECRv events with the energy of $10^{15}$eV to $10^{23}$eV. This energy level is in accordance with the discovery of the PeV neutrinos by IceCube \citep{icecube}, \citep{icecube2}, where the possibility of the existence of neutrinos with energy above 100 PeV is now being considered. The corresponding physics including the cross-section of this regime of neutrinos are discussed in the literature. (e.g. \citep{marfati})\\
We use frequencies of 150 MHz and 1.5 GHz as indicators of UHECRv observations in MHz and GHz bands. These are the same frequencies used in the reference papers \citep{ref01} and \citep{ref02}.  

Lunar UHECRv events at low frequencies also have been  investigated using numerical methods \citep{Stal}, \citep{Gusev}. \\ In our analysis, the bandwidth (B.W.) is roughly set to $13\%$ of central frequency ($f_{c}$) (20 MHz for $f_{c}$ of 150 MHz and 200 MHz for $f_{c}$ of 1.5 GHz). The antenna in this analysis is assumed to be a simple resonance antenna, which is optimised at its central frequency. Thus only a narrow bandwidth would allow for a widebeam and omnidirectional radiation pattern in this design. Using this antenna at a broader bandwidth requires additional components for impedance matching. Also at a broader frequency range, side-lobes appear in the radiation pattern, which makes the calibration complicated. \\ Coherent Askaryan radiation is the dominant radio emission generated by the UHECRv events at both frequency regimes. For a lunar lander antenna, surface waves are the main mechanism for propagation of the radiation of those events which occur at a far distance on the Moon's surface. At low frequencies, kHz up to a few MHz, transition radiation might be also important as the emission mechanism \citep{transitionkhz} for upcoming neutrinos interacting very near the lunar surface. \\ The selected frequency band at 150 MHz is within the optimum frequency window where the wavelength is of the order of  the length of the shower. As a result, it  produces a more isotropic UHECRv radio emission \citep {Scholten}. 
\subsection{Antenna and Receiver}
A tripole antenna is a preferred choice for individual antennas onboard the lunar orbiter satellites and lunar lander \citep{Stal,KleinW}. It consists of 3 co-centred orthogonal dipoles which enables it to detect signals in all directions. In principle, a single tripole antenna is capable of localising radio emissions with a few degrees resolution \citep{Chen} . Considering half-wavelength (resonance) dipoles as the optimal design, the length of dipoles would be around 1 m for  $f_{c}$ of 150 MHz and 10 cm for $f_{c}$ of 1.5 GHz which make them suitable for space-based experiments. In contrast, the large size of antenna at kHz frequencies could be an issue for lunar missions. The exact length and orientation of antenna should be optimised based on the lunar environment and system characteristics. \\
In our analysis for onboard antennas, the gain of the antenna is set to 1 so antennas are assumed to be omnidirectional. The corresponding $A_{e}$ in 
Eq.\ref{eq:1} then becomes  $A_{e}= {\lambda}^2 / 4\pi$ where $\lambda$ is the wavelength of the operational frequency. This assumption simplifies the calculation of Eq.\ref{eq:1} otherwise $A_{e}$ becomes a complicated function of the antenna gain and should be calculated for each frequency over the entire bandwidth.\\ 
The receiver sensitivity is identified with the minimum detectable electric field. We apply the formula (Eq.\ref{eq:1}) presented in \citep{glue} and \citep{ref01}. For the 150 MHz band, the dominant system noise in the selected B.W. is Galactic noise and it varies between $820^{\circ}$ K (for 100 MHz) and $142^{\circ}$ K (for 200 MHz)\citep{Tformula}. In the meantime, the Moon's noise temperature varies in a wide range from $130^{\circ}$ K to $400^{\circ}$ K \citep{moontemperature} which also means that only Extreme Temperature Electronics (ETE) can be operational on the Moon's surface. By using a Low Noise Amplifier (LNA) which typically has a noise figure below 1.5 (equivalent to noise temperature lower than $120^{\circ}$ K ) at the first input of the receiver, the dominant noise is usually Galactic noise at MHz regime. In our analysis, the Galactic noise temperature at central frequency $(294^{\circ}$ K, 150 MHz) has been used as the system noise. This is roughly equal to the noise temperature of a receiver with a noise figure of 3 dB. With noise temperature of $294^{\circ}$ K, the minimum detectable electric field for the receiver becomes 2.63 $\mu$V/m per MHz which can be improved by using more antennas. For 1.5 GHz the dominant noise temperature is generated by electronics and depends on system design and operating environment.  For our analysis at 1.5 GHz we used a conservative receiver temperature of $300^{\circ}$ K, roughly the same as amount of the dominant noise at 150 MHz. The corresponding minimum detectable electric field for a receiver at 1.5 GHz becomes 8.41 $\mu$V/m per MHz, which is more than three times at 150 MHz. Therefore only UHECRv events at higher radiation amplitude are expected to be detected at 1.5 GHz. As we will see this is reflected in the event rate predictions of the two frequency regimes. For ground-based observation, minimum detectable electric field is set at 0.01 $\mu$V/m per MHz, a typical number which has been used in \citep{ref01} and \citep{ref02}. For simplicity, we assume the bandwidth and Tsys of such an array to be the same as those for a lunar lander or lunar orbiter antenna. (Tsys=$294^{\circ}$ K and B.W.=20 MHz for 150 MHz, Tsys=$300^{\circ}$ K and B.W.=200 MHz for 1.5 GHz). This gives the collecting area of the array as 22100 square meters for 150 MHz and 2255 square meters for 1.5 GHz. 
It should be noted that the same sensitivity can be achieved with a smaller collecting areas if a broader B.W. is used. For instance, by using the B.W. of 100 MHz at a central frequency of 150 MHz, the collecting area can be reduced to 4420 square meters. For a frequency of 1.5 GHz and B.W. of 0.5 GHz, the collecting area of 902 square meters would be needed to reach the sensitivity of 0.01 $\mu$V/m per MHz. In tables \ref{tab:i} and \ref{tab:ii}, a summary of receiver parameters in GHz and MHz used in this study is shown. In the next section the analysis method and results for UHECRv events are discussed. 
\begin{table*}[t]
\centering
\begin{tabular} {l*{6}{c}r}
Radio Observation              & $f_{min}$ & $f_{max}$ & $B.W.$ & $f_{0}$ & $E_{min}$(V/m/MHz)  & Frequency Band \\
\hline
Lunar Lander & 1.4 & 1.6 & 0.2 & 1.5 & $8.41\times10^{-6}$ & GHz   \\
Lunar Orbiters& 1.4 & 1.6 & 0.2 & 1.5 & $8.41\times10^{-6}$ & GHz   \\
Ground-based array &1.4 & 1.6 & 0.2 & 1.5 & $1\times10^{-8}$  & GHz    \\
SKA1 MID2   & 0.950 &1.760 & 0.81 &1.355&  $2.1\times10^{-9}$  &GHz   \\

\end{tabular}
\caption{\label{tab:i} Receiver parameters of radio UHECRv observations in GHz band. Frequencies and bandwidths are in GHz. }
\end{table*}
\\
\\
\begin{table*}[t]
\centering
\begin{tabular}{l*{6}{c}r}
Radio Observation              & $f_{min}$ & $f_{max}$ & $B.W.$ & $f_{0}$ & $E_{min}$ (V/m/MHz)  & Frequency Band \\
\hline
Lunar Lander & 140 & 160 & 20 & 150 & $2.63\times10^{-6}$& MHz    \\
Lunar Orbiters & 140 & 160 & 20 & 150 & $2.63\times10^{-6}$& MHz    \\
Ground-based array   & 140 & 160 & 20 & 150 & $1\times10^{-8}$& MHz    \\
SKA1 LOW   &50 & 350 &300 & 200 &$1.4\times10^{-9}$  &MHz   \\ 
LOFAR &110& 130 & 20 & 120 & $6.62\times10^{-9}$ &MHz    \\
\end{tabular}
\caption{\label{tab:ii} Receiver parameters of radio UHECRv observations in MHz band. Frequencies and bandwidths are in MHz. $E_{min}$ LOFAR is calculated from the array parameters in \cite{Scholten}. }
\end{table*}

\begin{figure*}[t]
\centering 
\includegraphics[width=0.8\textwidth]{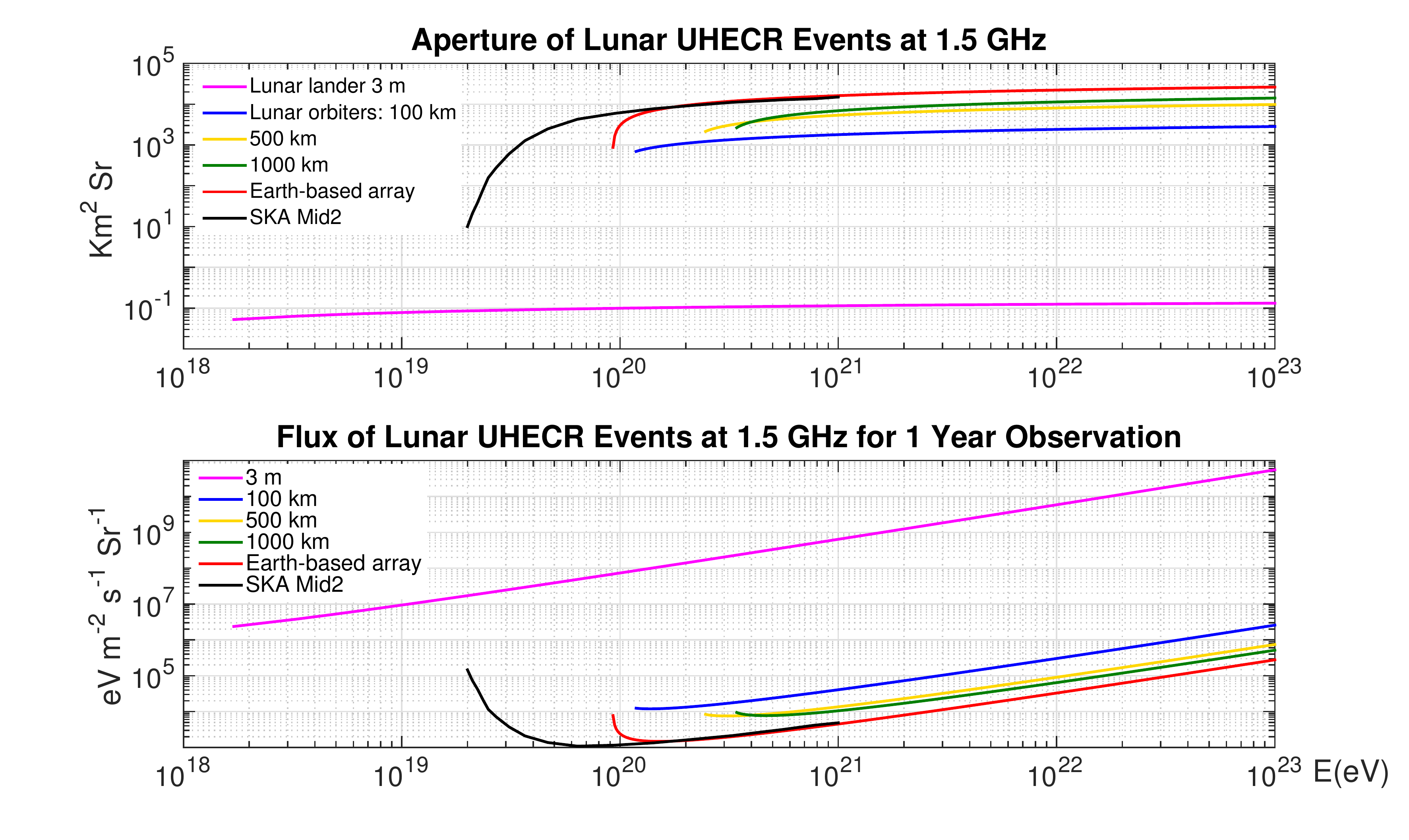} 
\hfill
\caption{\label{fig:iii} Top, Apertures of UHE Cosmic Rays for individual antenna onboard  a lunar lander 3 m above Moon's surface and onboard lunar orbiters at altitudes of 100 km, 500 km, 1000 km with a sensitivity of 8.41 $\mu$V/m/MHz, also for ground-based array with the sensitivity of 0.01 $\mu$V/m/MHz. Results are compared with the aperture of SKA Mid2 (reproduced from \citep{ska}). See table \ref{tab:i} for receiver parameters. Bottom, corresponding flux limits of UHE Cosmic Rays for a 1-year observation at 1.5 GHz.}
\end{figure*}

\begin{figure*}[t]
\centering 
\includegraphics[width=0.8\textwidth]{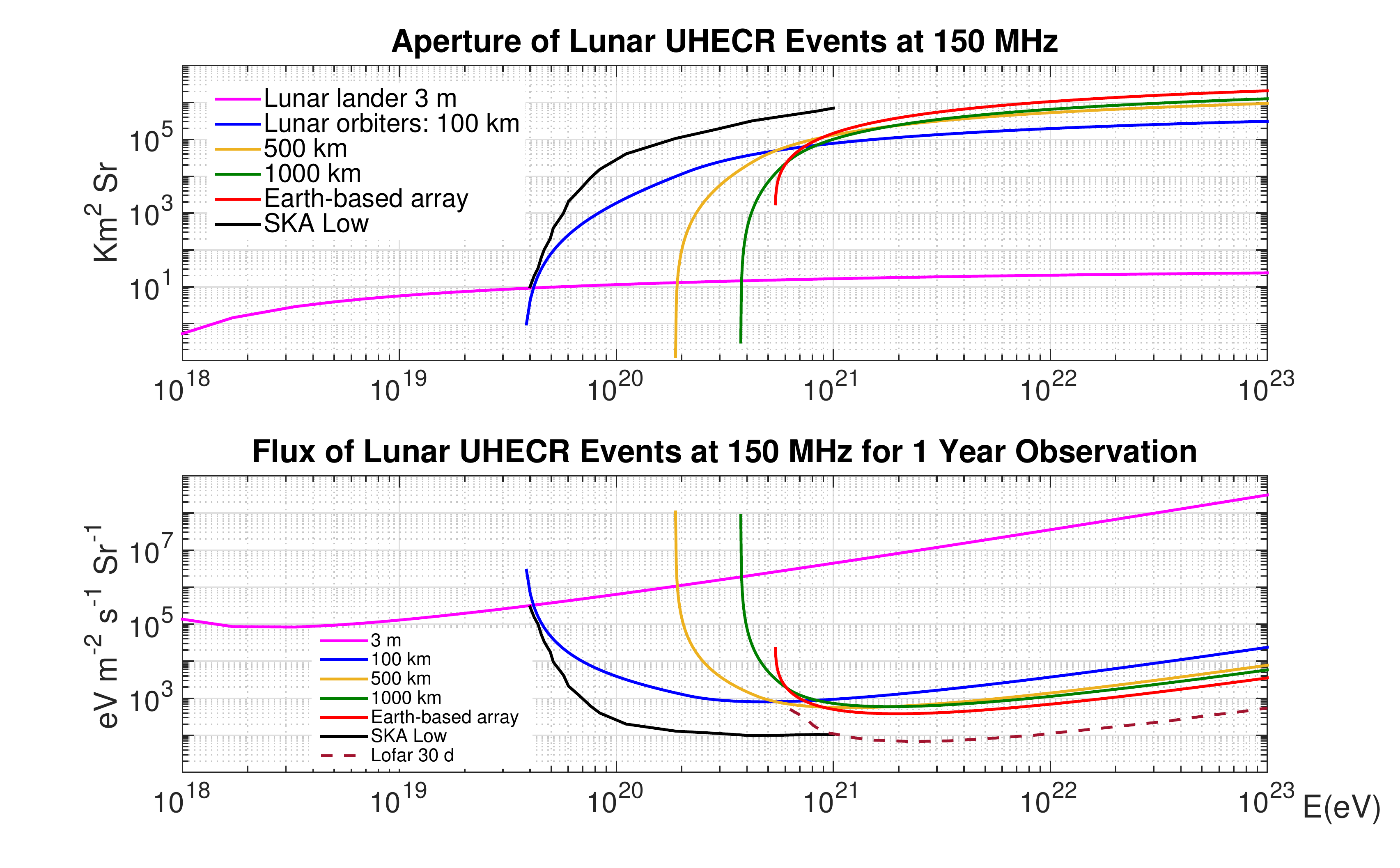} 
\hfill
\caption{\label{fig:iv} Top, Apertures of UHE Cosmic Rays for individual antenna onboard  a lunar lander 3 m above Moon's surface and onboard lunar orbiters at altitudes of 100 km, 500 km, 1000 km with a sensitivity of 2.63 $\mu$V/m/MHz, also for ground-based array with a sensitivity of 0.01 $\mu$V/m/MHz. Results are compared with the aperture of SKA Low (reproduced from \citep{ska}). See table \ref{tab:ii} for receiver parameters. Bottom, corresponding flux limits of UHE Cosmic Rays for a 1-year observation at 150 MHz. Results are compared with flux limit of LOFAR for a 30-day observation (reproduced from \citep{Scholten}). }
\end{figure*}

\begin{figure*}[t]
\centering 
\includegraphics[width=0.7\textwidth]{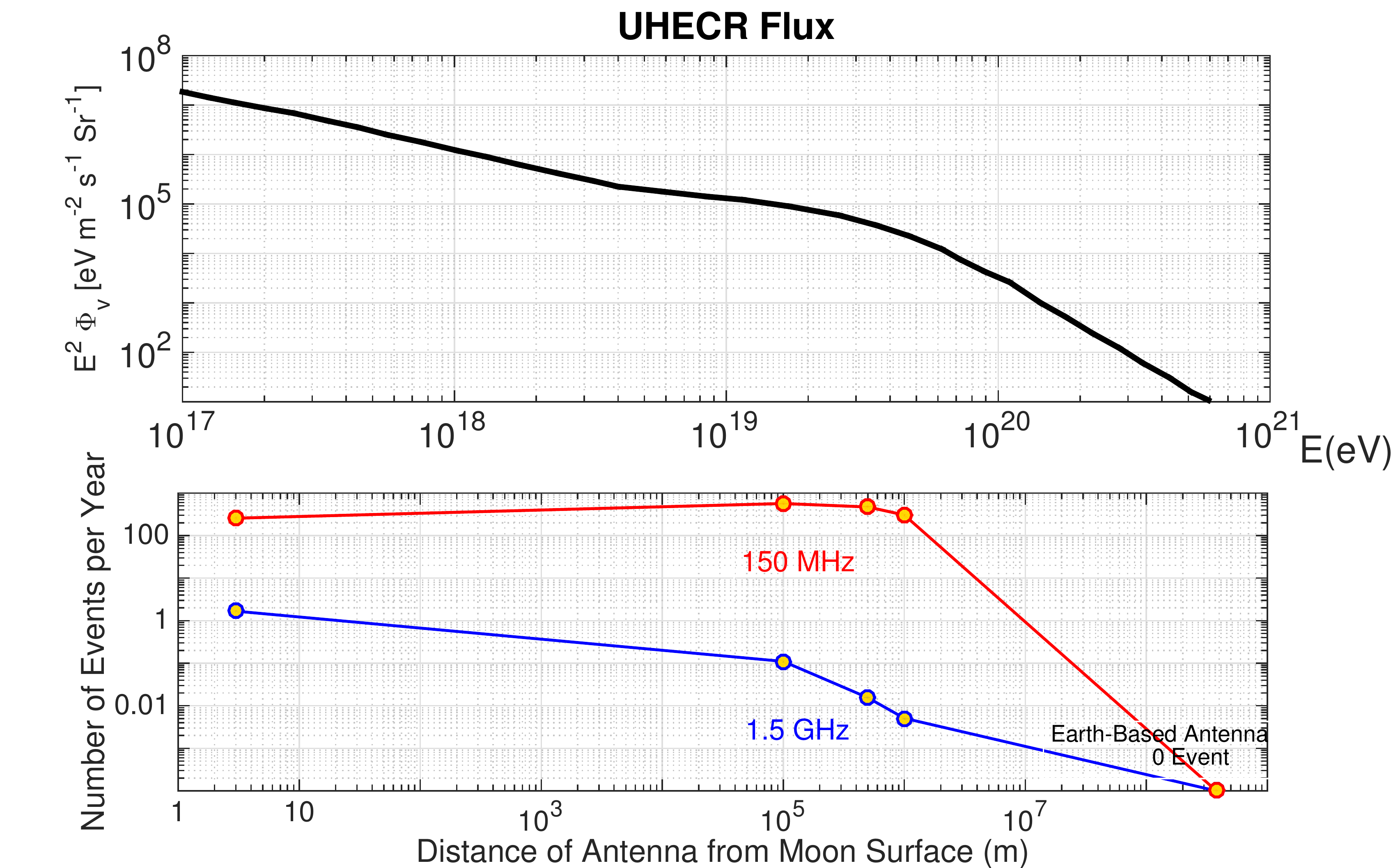} 
\hfill
\caption{\label{fig:v} Top, predicted UHECR flux model (reproduced from \citep{CRflux}. Bottom, prediction of number of events for 1-year observations of UHE Cosmic Rays for antennas
onboard  a lunar lander 3 m above the Moon's surface and onboard lunar orbiter satellites at altitudes of 100 km, 500 km, 1000 km and Earth-based antenna array (all marked with yellow circles). Results are shown for both frequencies of 150 MHz (Red) and 1.5 GHz(Blue).}
\end{figure*}

\begin{figure*}[t]
\centering 
\includegraphics[width=0.8\textwidth]{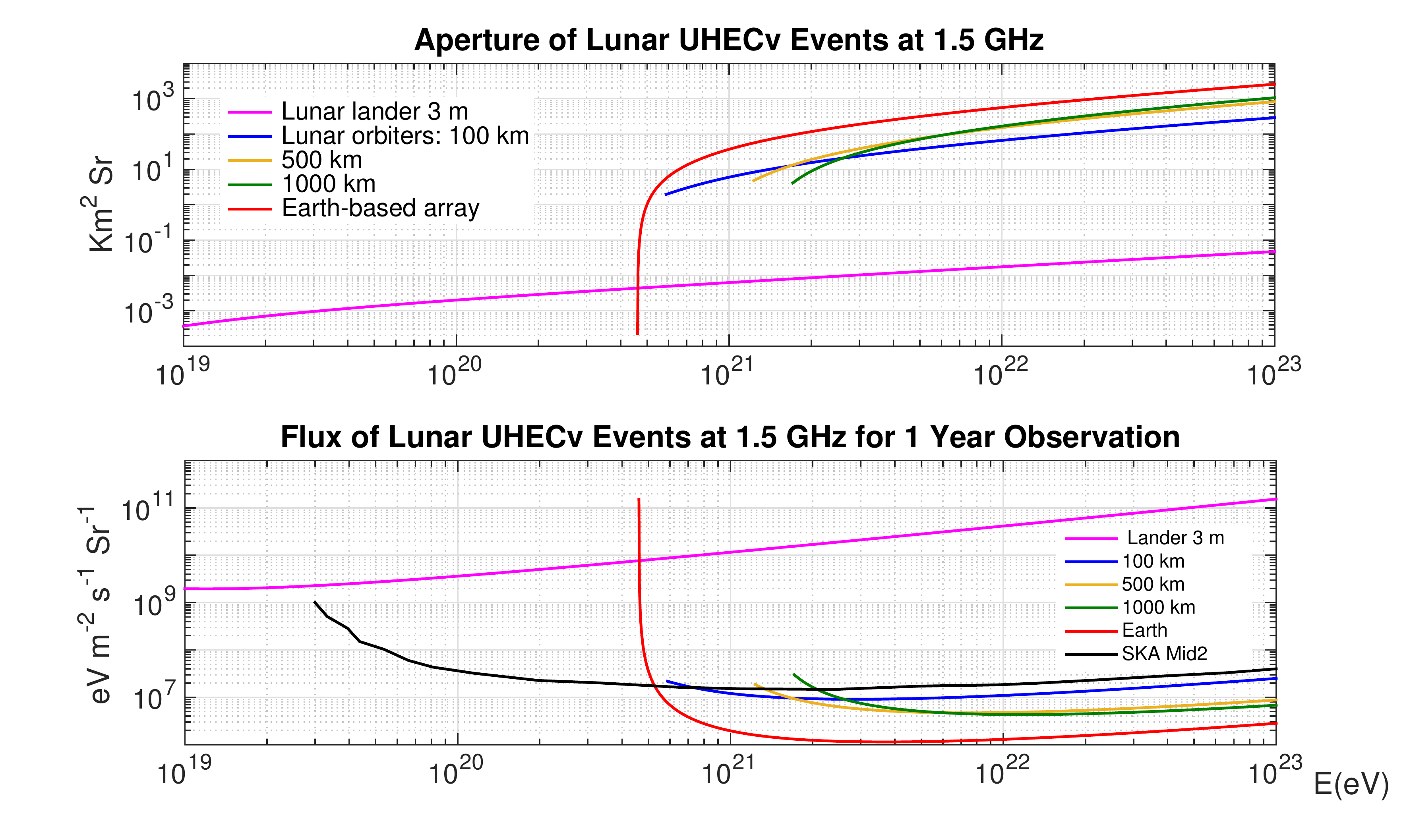} 
\hfill
\caption{\label{fig:vi} Top, Apertures of UHE Cosmic Neutrinos for individual antenna onboard a lunar
lander 3 m above Moon's surface and onboard lunar orbiters at altitudes of 100 km, 500 km,
1000 km with a sensitivity of 8.41 $\mu$V/m/MHz, also for ground-based array with a sensitivity of 0.01 $\mu$V/m/MHz. Bottom, corresponding flux limits of UHE Cosmic Neutrinos for a 1-year observation at 1.5 GHz. Results are compared with the aperture of SKA Mid2 for a 1000-hour observation(reproduced from \citep{ska}). (See table \ref{tab:i} for receiver parameters).} 
\end{figure*}

\begin{figure*}[t]
\centering 
\includegraphics[width=0.8\textwidth]{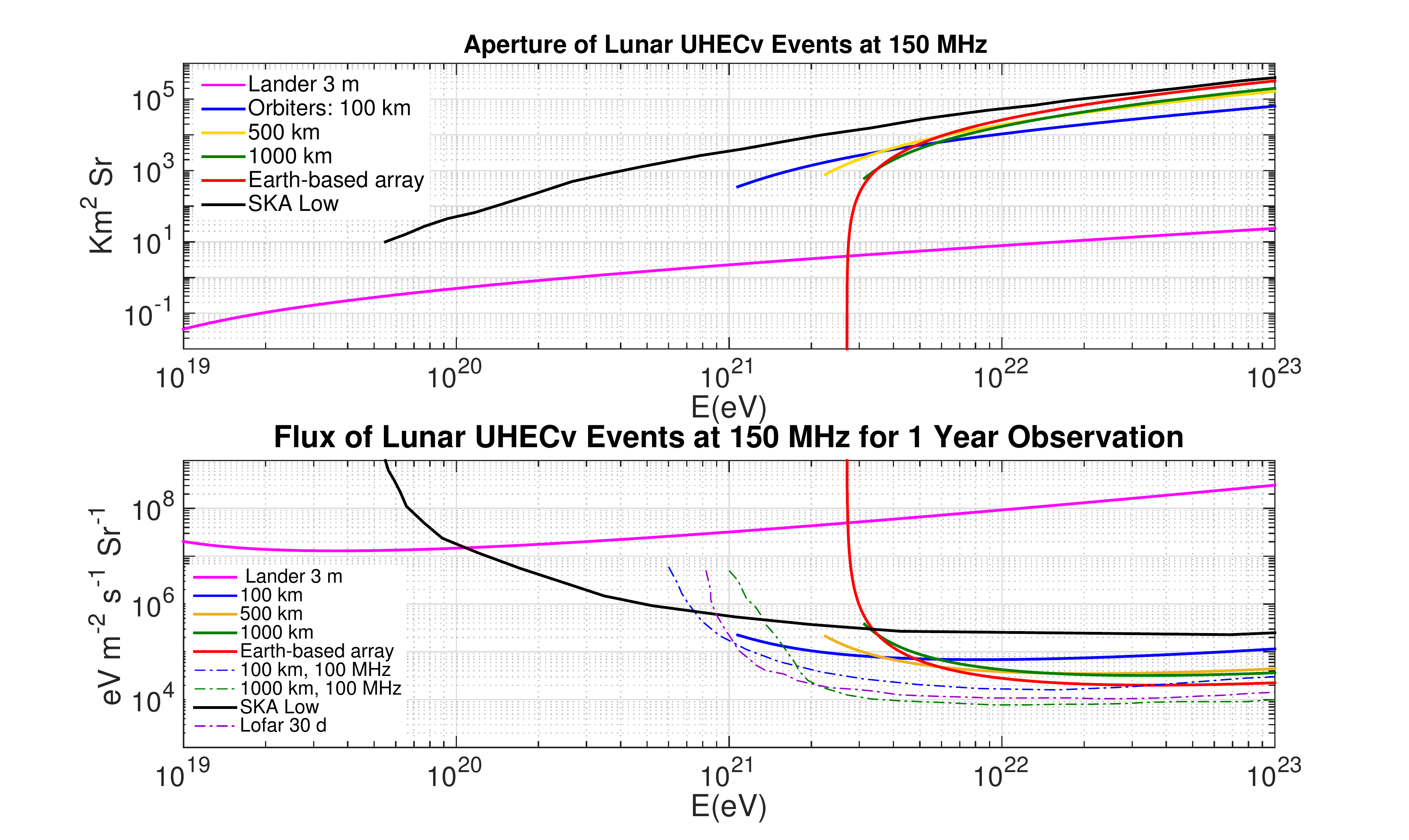} 
\hfill
\caption{\label{fig:vii} Top, Apertures of UHE Cosmic Neutrinos for individual antenna onboard a lunar
lander 3 m above Moon's surface and onboard lunar orbiters at altitudes of 100 km, 500 km,
1000 km with a sensitivity of 8.41 $\mu$V/m/MHz, also for ground-based array with a sensitivity of 0.01 $\mu$V/m/MHz and SKA Low \citep{ska}). Bottom, corresponding flux limits of UHE Cosmic Neutrinos for a 1-year observation at 150 MHz. Results are compared with aperture of SKA Low for a 1000-hour observation(reproduced from \citep{ska}), LOFAR, 30 days observation \citep{Scholten}, individual antenna at 100 MHz at 100 km and 1000 km lunar orbiter for 1 year obaervation \citep{Stal}. (See table \ref{tab:ii} for receiver parameters)}
\end{figure*}

\begin{figure*}[t]
\centering
\includegraphics[width=0.7\textwidth]{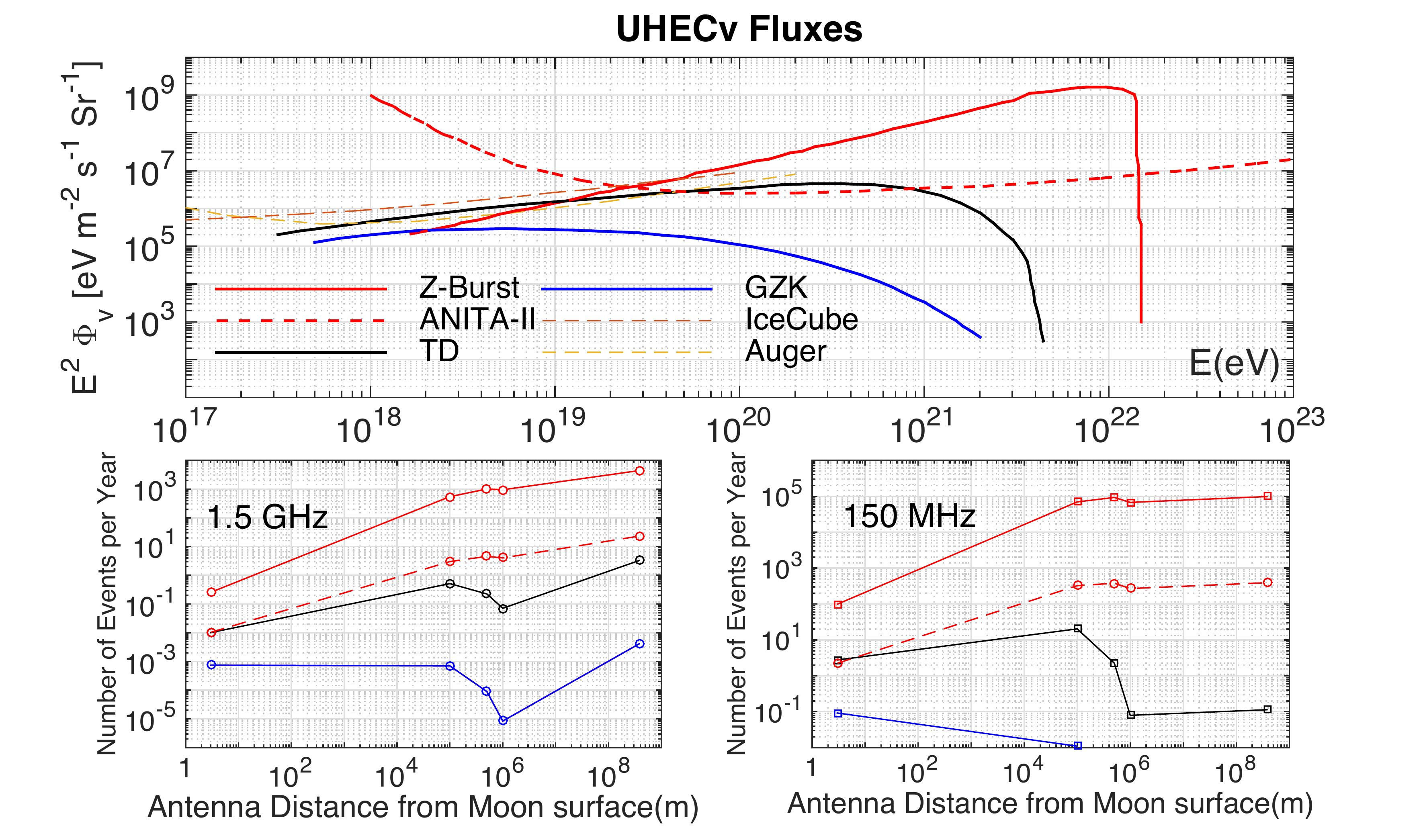}
\hfill
\caption{\label{fig:viii} Top, Predicted fluxes for GZK  neutrinos \citep{GZK}, Topological Defects (TD) and Z-Burst processes \citep{TDZ}. Also limits of recent neutrino observations: ANITA-II \citep{ANITA2}, Auger \citep{auger-rej} and IceCube \citep{icecube}.  Bottom, the expected events for a 1-year observation of lunar UHECv events. Event rates are for observations at 1.5 GHz (left) and 150 MHz(right) and are plotted vs.\ distance of the antenna from the Moon.} 
\end{figure*}
\section{Analysis of Lunar UHECRv Events}
\label{secap}
\subsection{Lunar Cosmic Ray Events}
We use the method developed in \citep{ref02} for detection of UHE Cosmic Rays colliding with the Moon. Cosmic Rays, containing energetic primary particles, unlike neutrinos, can not penetrate through the lunar regolith, therefore, CR events only occur on the lunar surface. The radio emission then propagates mostly through the lunar exosphere where radio wave attenuation can be neglected. The calculation is based on UHECR aperture which is the physical aperture illuminated by the antenna times a probability function. Here we show it as $P_{CR}(E)$ and it is the probability of CR interaction with the Moon's surface which produces Askaryan radiation in the aperture. $P_{CR}(E)$ depends on the frequency of observation, Askaryan radiation characteristics and lunar environment properties including refractive index. It also includes the contribution from smooth surface and surface roughness.
The formula for UHECR aperture is presented in \citep{ref02} and we use it for different experiments as follows. For the lunar lander, the antenna is assumed to be 3 m above the lunar surface which was the height of the ESA lunar lander planned for LRX. For lunar orbiters, results for an individual antenna at altitudes of 100, 500 and 1000 km from the lunar surface are shown. Also, results for ground-based arrays are shown for two frequency regimes. We use the same sentivity taken from Eq.\ref{eq:1} for individual antennas but the field of view (FoV), and as a result the area of the Moon's surface covered by a lunar lander antenna or the lunar orbiter antennas, depends on the distance of antenna from the Moon's surface. In the next section the corresponding aperture for each case is calculated. 
\subsubsection{ Aperture, Flux Density and Event Rate of UHECR Events}
UHECR Apertures at 1.5 GHz vs.\ energy are illustrated in the top plot of Fig.\ref{fig:iii}. For all results the apertures are a small portion of the actual aperture of Moon's surface illuminated by antenna:



\begin{eqnarray}
\label{eq:2}
\hspace*{-0.7cm} Ap= A_{0} \, . \, P_{CR}(E) (km^2 .Sr), \quad \quad
\hspace*{-0.7cm} A_{0}=2 \pi \, . \, \pi R_{0}^2 (km^2 .Sr)
\end{eqnarray}

Where $R_{0}$ is the radius of lunar surface which is covered by antenna radiation pattern. An omnidirectional pattern is assumed for antennas on the Moon's surface. For lunar orbiter and ground based antennas, it is assumed that the surface is illuminated with the antenna main-lobe so the gain is constant over the whole beamwidth. Under these conditions $R_{0}$ depends on the distance of antenna from the lunar surface and can be calculated using simple geometry. Maximum $R_{0}$, $R_{ap}$, for various distances are included in table \ref{tab3}. Detectable UHECR events can occur within a radius of $R_{0}$ from the point where antenna is vertically mapped to the lunar surface. Therefore the total aperture is integrated over $A_{0}(r) . P_{CR} (E,r) $ where r varies between 0 and $R_{ap}$. (See \ref{A}) 

\begin{table}[t]
\centering
\begin{tabular}{c|c}
Distance from Moon & $R_{ap}$(km) \\
\hline
Lunar Lander: 3 m&3.23 \\
Lunar Orbiters: 100 km &565.46 \\
500 km& 1095\\
1000 km & 1343\\
Ground-based array: 380000 km & 1738 (Moon's Rad.) \\
\end{tabular}


\caption{\label{tab3} Radius ($R_{ap}$) of physical area on the Moon's surface in the field of view of omnidirectional antennas at various distances from Moon.}
\end{table}

Although physical aperture increases at larger distances, Askaryan radiation at the position of the antenna becomes weaker due to the distance. As a result, the detection of UHECR events is a compromise between the distance of antenna and size of aperture. This has been reflected in simulated results in Fig.\ref{fig:iii}.  The apertures are shown via energy of cosmic rays in a broad range from $10^{18}$ eV to $10^{23}$ eV. The possibility of the detection of UHECRv events increases with the primary energy of events, however the detectable aperture is limited by the physical aperture of the Moon's surface which is illuminated by the antenna. This is particularly the case for antenna at 3 m height whose aperture is almost constant over a broad range of energy (Fig.\ref{fig:iii} and Fig.\ref{fig:iv}). As distance increases, lower energy events become undetectable but the size of aperture increases significantly. At energies of $3 \times 10^{20}$eV and higher the aperture reaches a threshold as the radius of aperture becomes comparable to the lunar radius. Therefore antenna on lunar orbiters at order of 500 km -1000 km would be the optimum distance for detecting lunar UHECR events. In Fig.\ref{fig:iii} apertures of ground-based observation and SKA Mid2 (190 SKA antennas) are also shown for comparison.
In the bottom plot, the corresponding flux limits of the above experiments are shown. The illustrated flux results ($E_{cr}^2  \Phi_{cr}$) represent the detectable flux densities of UHECR events for a 1-year observation for $90\%$ confidence limits (e.g. \citep{Stal}).\\
The same method has been used to extract apertures for UHECR events at 150 MHz. Results are illustrated in Fig.\ref{fig:iv}. Comparing with Fig.\ref{fig:iii}, it can be seen that aperture at higher energy levels is proportional to the square of wavelength. ($A_{p}\propto \lambda^2$). The curves of detectable CR events are slightly shifted towards the lower energy levels at GHz band. For instance, this is seen for the aperture of a ground-based array with the same sensitivity where at 150 MHz the detectable events begin around $6 \times 10^{20}$eV compared with $10^{20}$ eV for 1.5 GHz. Similar to the apertures at 1.5 GHz,
there are small changes in the size of the aperture at distances higher than 500 km which makes it an optimum distance for radio observations of lunar UHECR events at 150 MHz. \\ In the bottom plot of Fig.\ref{fig:iv}, corresponding flux limits of various experiments are shown and simulated results are compared with the expected outcome for 30 days of UHECR observation with LOFAR. In both frequency regimes, the events with energy lower than $10^{19}$ eV can only be detected by antenna on the Moon's surface.\\
Using calculated apertures and standard flux densities of UHECR, it is possible to estimate the event rate of each experiment. Here we use predicted flux for the occurrence of UHECR presented in \citep{CRflux}. The model and expected event rates for various lunar observations are shown in Fig.\ref{fig:v}.  The model is for UHECR with energy level between $10^{17}$ and $10^{21}$ eV. At this range of energy of UHECR and using calculated apertures, expected detectable event rate are calculated. The results are shown for two frequency regimes (1.5 GHz in blue and 150 MHz in red) for a 1-year observation. By comparing the number of expected events per year, it can be seen that no UHECR events using Earth-based observation with ($E_{min}=0.01\mu$V/m/MHz) within this window of energy would be detected. Also from the UHECR flux in Fig.\ref{fig:v} one would expect that most of the expected events has an energy in the range of $10^{17}$ to $10^{18}$ eV where the flux density is high. The lunar lander antenna becomes particularly important for detection of lunar cosmic ray events at lower energy range where other experiments can not detect these events due to the weak signal. We will discuss this in details in \ref{lowenergy}. 
 
\subsection{Lunar Cosmic Neutrino Events}
We modified method in \citep{ref01}  to analyse lunar neutrino events (UHECv) for various experiments. Here are the main assumptions:\\
- $E_{s}=0.2 . E$\\
For neutrino events only 20\% of energy of neutrinos is converted to the hadronic shower while $E_{s}=E$ for Cosmic Rays (e.g.\citep{bray}).\\
Similar to Cosmic Rays, an aperture is defined based on the physical aperture covered by antenna radiation pattern and a possibility of the occurrence of UHECv events:


\begin{equation}
\label{eq:3}
Ap= A_{0} \, . \, P_{v}(E) (km^2 .Sr),
A_{0}=4 \pi \, . \, \pi R_{0}^2 (km^2 .Sr) 
\end{equation}

While for UHECR only downward CR are taking into account, both upward  and quasi-horizontal downward  
neutrinos are effective so a full sphere ($4\pi$) solid angle is applied for calculation of UHECv apertures.  Similar to cosmic rays, contribution from surface roughness is included in $P_{v}(E)$ as it scatters the radiation for surface events. Results of analysis show that at selected frequencies the major contribution is from downward neutrinos. This is consistent with results of ground-based observations in \citep{ref01}. More recent studies suggest that small-scale surface roughness could have greater contribution in some UHECRv events but on average, the contribution of surface roughness would be small \citep{jamesrough}. \\
For an antenna 3 m above the lunar surface, we add the possibility of UHEv events occurring  in the lunar sub-regolith. This is done using the formula presented in \citep{fdtd3} and by taking into account the attenuation of radio wave propagation through the lunar regolith. Typical numbers for refractive index of lunar regolith ($n_r$=1.73) and sub-regolith ($n_r$=2.5) \citep{fdtd3}, \citep{fdtd5} have been used in the aperture calculation. This corresponds to dielectric constant ($\epsilon_{r}$) of 3  and 6.25 respectively. For attenuation, the loss tangent of 0.01 (tan($\delta) = 0.01)$ \citep{losstangent} is set and the depth of regolith is assumed to be 10 m \citep{regolith10m}.  We chose loss tangent and depth of regolith as extreme cases to estimate the maximum attenuation of Askaryan radiation when passing through the lunar regolith. The attenuation reduces the maximum electric field of Askaryan radiation of sub-regolith. The sub-regolith aperture is calculated similar to the regolith aperture but the roughness effect has been eliminated. This is because that sub-regolith radiation could face multiple internal reflections before reaching the surface therefore the effect of roughness over the whole aperture becomes complicated. The calculation for sub-regolith, therefore, represent lower limits to the apertures but as discussed, the roughness effect has a small contibution in calculation of the apertures. The total aperture for UHEC neutrinos then becomes a virtual sphere which covers both events on the lunar surface and those which occur inside the lunar regolith. The details of calculation are explained in \ref{A}. In this analysis only the attenuation of the direct path between lunar sub-regolith and antenna is taken into account. However, the radiation generated in a cascade shower could in principle face multiple reflections through sublayers and the antenna might receive a superposition of radio emission in various directions. This could be particularly significant for distant lunar observations where contributions from multiple cascades of Askaryan showers result in a global radiation, which would be different from individual local radio emission. This is the case for the effect of sub-layers of the Moon regolith for both lunar orbiter and ground based experiments. The analysis requires further investigations and we leave it for future study. 
\subsubsection{ Aperture, Flux Density and Event Rate of UHECv Events}
Apertures of lunar UHECv events at 1.5 GHz for various distances are shown in Fig.\ref{fig:vi}. Compared with UHECR apertures at the same frequency, UHECv apertures have been roughly decreased  by the order of 100, but it follows a similar trend. For an antenna 3 m above the lunar surface and sensitivity of 8.41 $\mu$V/m/MHz, aperture covers the entire range of energy from $10^{19}$eV to $10^{23}$eV. 
As opposed to CR aperture, the aperture increases almost linearly from about 0.001 $km^2Sr$ \ at\ the\ energy\ of\ $10^{19}$eV to 
$10^{-1} km^2Sr$ \ at\ the\ energy\ of\ $10^{23}$eV.
For 100 km distance, detectable events begin at $6 \times 10^{20}$eV. As the orbiter distance increases, only energetic UHECv events at higher levels become detectable so that for 1000 km the detectable events begin around $2 \times 10^{21}$eV. For Earth-based arrays with a sensitivity of 0.01 $\mu$V/m/MHz, detection of UHECv events begins at energy of $5 \times 10^{20}$eV. Similar to UHECR apertures, lunar orbiter experiments at 500 km and 1000 km distance approach a constant level at the highest energy level. As a result, it seems that this range of altitudes is an optimum distance for lunar observations of UHECRv events. In the bottom plot of Fig.\ref{fig:vi}, corresponding flux limits of UHECv apertures for a 1-year observation are plotted. For comparison, the flux limit of SKA Mid2 (see table \ref{tab:i} ) for 1000 hours of observations is also plotted \citep{ska}. Similarly,  aperture and fluxes of lunar UHECv events at 150 MHz vs.\ distance are shown in Fig.\ref{fig:vii}. Similar to CR, neutrino apertures are proportional to the square of wavelength at higher energy levels ($A_{p}\propto \lambda^2$). The minimum energy level of detectable neutrino events are also slightly shifted as the frequency changes. For instance, UHECv events at 100 km become detectable at $10^{21}$eV at 150 MHz compared with $5\times10^{20}$eV at 1.5 GHz (the receiver parameters are available in Tab.\ref{tab:i} , Tab.\ref{tab:ii}
). For Earth-based array, this is also the case where UHECv events can be observed from $3 \times 10^{21}$eV at 150 MHz and  $4\times10^{20}$eV at 1.5 GHz (with array sensitivity of 0.01 $\mu$V/m/MHz). The maximum aperture at highest energy levels approaches a constant level for altitudes of 500 km and 1000 km. The aperture of SKA Low is illustrated \citep{ska} in the plot for comparison. In the bottom plot of Fig.\ref{fig:vii} , corresponding flux limits of UHECv events at 150 MHz for a 1-year observation are plotted. Flux limits of LOFAR (30 days) \citep{Scholten} and SKA Low (1000 hours) \citep{ska} observation are shown in the plot. It should be noted that data for SKA array only covers the antennas at phase I. For instance, the SKA2 MID2 will use  1,500 SKA antenna dishes \citep{ska}.
Also the flux limit for an individual antenna at 100 MHz for 1-year observation at 100 km and 1000 km on a lunar orbiter is shown 
\citep{Stal}. The results of \citep{Stal} have been calculated using a numerical Monte Carlo method and the antenna parameters are different. As is expected, however, the aperture increases with the wavelength,  and as a result the flux of UHECv events at 150 MHz approaches a lower limit compared with the one for 100 MHz. \\
Similar to \citep{Stal}, the expected lunar UHECv event rate for 3 models of neutrino process is presented in Fig.\ref{fig:viii}. Neutrino flux models are reproduced on the top plot.  The Greisen-Zatsepin-Kuzmin (GZK) model predicts GZK neutrinos which have a limit of about $5\times10^{19}$ eV. GZK neutrinos are generated during the UHECR interaction with Cosmic Microwave Background (CMB) via the secondary pion decay process. The detection of neutrinos at the GZK limit are important because it justifies the observations of UHECR with energies above the GZK limit \citep{GZK}. Topological Defects (TD) (e.g. \citep{Bhatta}) and Z-Burst processes (e.g. \citep{Weiler} ) are the suggested models for the existence of neutrinos with energies well beyond the GZK limit. The limits of recent neutrino observations are illustrated in 
Fig.\ref{fig:viii}. Z-Burst model, however, is ruled out by recent observations therefore for calculation of the occurrence of Z-Burst neutrinos, we apply the limit of ANITA-II (ANtarctic Impulse Transient Antenna) observations \citep{ANITA2}. The ANITA-lite v flux limit  \citep{Barwick} is well above the ANITA-II limit and is not shown in the figure. Also, the flux density models are not affected by the limits of IceCube \citep{icecube} and Auger \citep{auger-rej} experiments. Based on the models, UHECv event rates are calculated for 1.5 GHz and 150 MHz observation. Results are illustrated vs.\ distance of antenna to the lunar surface. For 1-year observation at 1.5 GHz (bottom left), observations of GZK and TD neutrinos are unlikely for all experiments. Considering ANITA-II limit, Z-burst neutrinos (the most energetic events) are only expected to be detected by a lunar orbiter antenna and an Earth-based array.\\
For 1-year observations at 150 MHz (bottom right), detection of GZK neutrinos is not expected  for all experiments. However, for an antenna 3 m above the lunar surface the event rate is 0.1. Thus an array of a hundred antennas on the lunar surface, for instance, would provide an ideal experiment for detection of this interesting regime of neutrinos. TD neutrinos are only detectable for low distance experiments from the Moon (3 m, 100 km and 500 km). With applying the limit of ANITA-II observations to Z-Burst model, several hundred of lunar UHECv events are predicted for one-year observation of a lunar orbiter antenna or an Earth-based array. Overall, it seems that 150 MHz would provide a reliable window for radio observation of lunar UHECRv events. For Earth-based observations GHz band could be preferred as the effects of the Earth's ionosphere such as dispersion becomes less important \citep{ghzdispersion}. It should be noted that the distances from few meters above the Moon's surface up to few tens of kilometer (which is the typical altitude of lunar orbiter satellites) are not realizable for lunar UHECRv experiments. 


\begin{figure*}[tbp]
\includegraphics[width=1\textwidth]{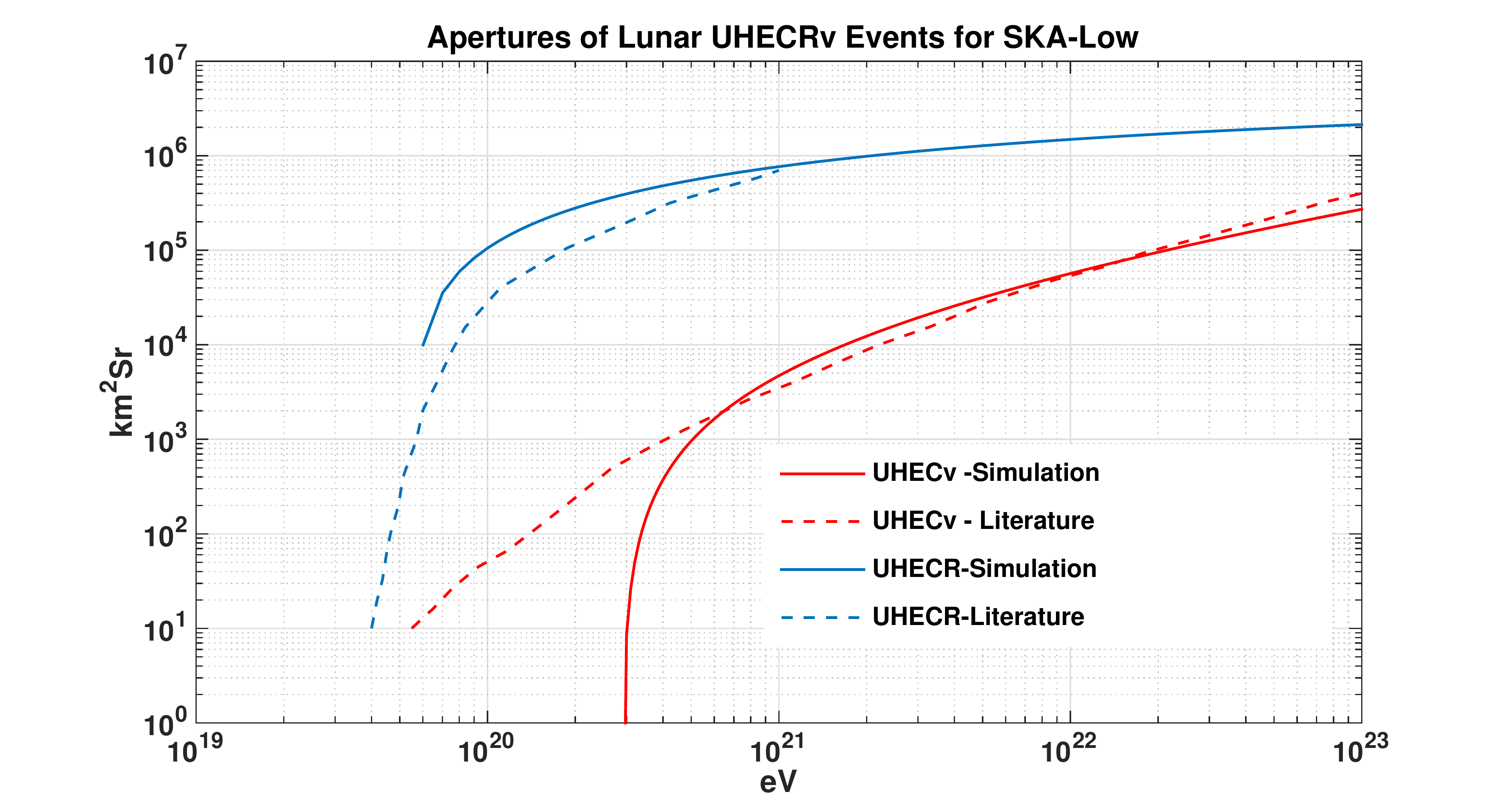} 
\hfill 
\caption{ Simulation of lunar UHECRv apertures (solid curves) for SKA-Low using parameters in table \ref{tab:ii}. Results are compared with the existing models (dashed curves) in \citep{ska}. }
\label{fig:skacomp}
\end{figure*}

\subsection{Validation of the Method}
To examine the validity of the analysis, a comparison of the method used in this article has been made with the existing models to calculate the apertures of lunar UHECRv events for SKA-Low. Results are shown in Fig. \ref{fig:skacomp} where curves on top (in blue) are the lunar UHECR apertures predicted for SKA-Low. The dashed curve in blue is taken from \citep{ska} which shows a max of 30\% difference with the results of our simulation at $10^{20}$ eV. Between $10^{20}$ eV and $10^{21}$ eV the difference between the two curves decreases, and at $10^{21}$ eV two curves approach the same point at $8\times10^{5}$ km Sr. For energies higher than $10^{21}$ eV the data for the existing model was not available. The lower curves (in red) compare two models for detection of lunar neutrinos using SKA-Low parameters. It can be seen that the calculated lunar UHECv aperture is in good agreement with the existing model \citep{ska} between $3\times10^{20}$ eV and $1\times10^{23}$ eV. The difference is that the UHECv events are not detectable at energy levels lower than $3\times10^{20}$ eV according to our analysis, while the energy limit for the existing model is $5\times10^{19}$ eV. Possible explanation is that a full Monte Carlo simulation is used in \citep{ska} where includes differential cross-sections in which sometimes 100\% of the neutrino energy would be converted to a hadronic cascade as opposed to 20\% for our analysis. It also should be noted that the detectable aperture becomes very small and non-practical at the level of $5\times10^{19}$ eV in our study.
\subsubsection{The Angular Spread of the Observations}
For all observations, a spherical symmetry is assumed so that the UHECRv events are detected randomly without specifying any preference over the viewing angle. In our analytical method, only the hadronic showers caused by a CR/ first interaction in the lunar regolith are considered. In principle, UHECRv events could also be detected by the second interactions. Examples are an upcoming UHECv event interacting in the regolith, with a secondary muon or tau. Also an initial neutral current interaction followed by a second interaction which could generate a secondary shower. Such a secondary shower is defined in the Monte Carlo simulation used in ANITA experiment \citep{Anita1}. It, however, should be noted that the expected apertures only increase if the first UHECRv interaction is out of the antenna view. This scenario is ruled out in our analysis since in both lunar orbiter and lunar lander observations, the antenna is assumed to be an omnidirectional antenna with a wide-beam radiation pattern.

\section{{Categorization of Lunar UHECRv Events }}
The categorization of UHECRv events presented in this study is aimed at pre-data processing and event triggering. This is particularly important for lunar missions where data reduction is unavoidable due to the limitations on the data transmission to the Earth. The full analysis of events including particle composition and energy of events will be done in post-processing and offline data analysis when data is received on Earth. For onboard processing, digital beamforming and direction of arrival (DoA) techniques are used for detection of the events.  
\subsection{Low Energy/ High Energy Events}
\label{lowenergy}
\begin{figure*}[t]
\includegraphics[width=1\textwidth]{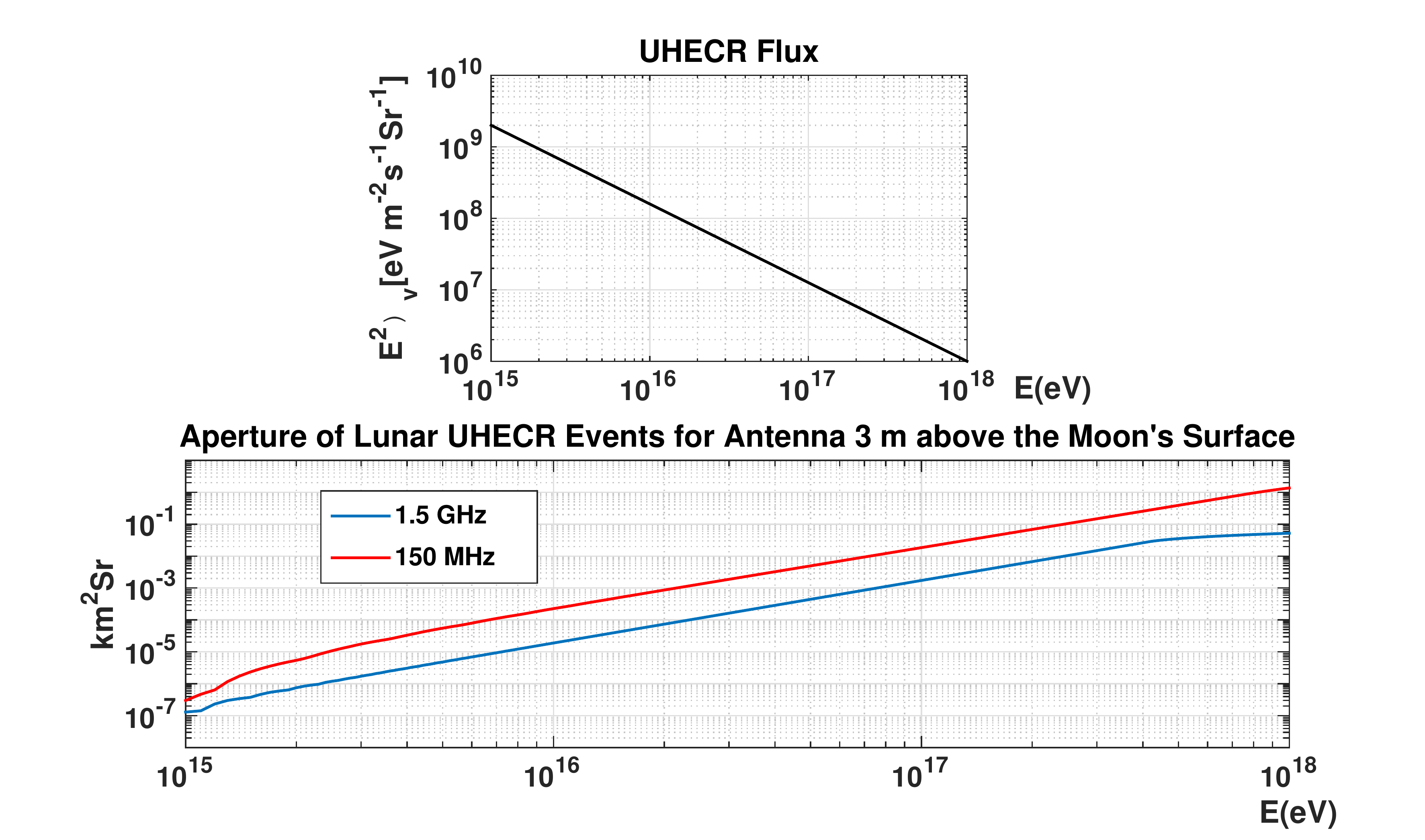} 
\hfill 
\caption{ Lunar CR apertures at lower energies of $10^{15}$ eV to $10^{18}$ eV for antenna 3 m above the Moon's surface. Apertures are calculated for frequency of 1.5 GHz (B.W. of 500 MHz) and frequency of 150 MHz (B.W. 50 MHz). The CR flux \citep{aj} is estimated between $10^{15}$ eV to $10^{18}$ eV. Using one antenna, the detection of 3 (1.5 GHz) and 5 (150 MHz) CR events per year is expected for this range of energy. }
\label{lowenergy3}
\end{figure*}

An estimate of CR flux between $10^{15}$ eV to $10^{18}$ eV is calculated from the CR flux spectrum \citep{aj}. This energy range in the spectrum is important because it is located between the regions known as Knee and Ankle and represents the transition between galactic and extragalactic cosmic rays. Due to the relatively low energy level of cosmic rays at this window, the only antenna on the lunar lander would be able to detect these events. We extend the calculation  for apertures of lunar lander antenna  at this region for both frequencies of 150 MHz and 1.5 GHz. It is found that with B.W. of 500 MHz at $f_{o}$ = 1.5 GHz and B.W. of 50 MHz at $f_{o}$ = 150 MHz, detection of 3 and 5 lunar CR events per year would be expected at this region. Corresponding flux densities and apertures are illustrated in Fig.\ref{lowenergy3}. Although this is an interesting science case for a lunar lander experiment, still distinguishing these low energy events remains unsolved. In the next section, the propagation of the Askaryan radiation in the lunar regolith is studied.


\begin{figure}[tbp]
\centering 
\includegraphics[width=.5\textwidth]{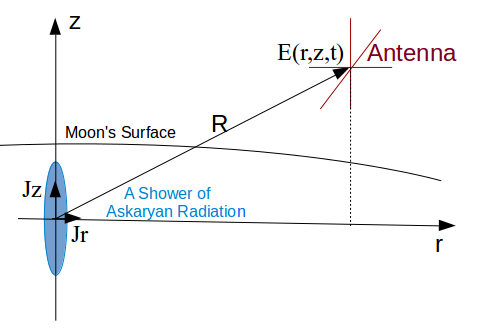} 
\hfill
\caption{\label{propag} An illustration of the analysis of Askaryan radiation in the lunar regolith. Jz and Jr are the vertical and radial components of the current density.  The current is due to the peak radio emission generated at the impact point of the energetic particles. E(r,z,t) represents the variation of the electric field in space and time. }
\end{figure}

\subsubsection{Electric Fields of Askaryan Radiation for Cascade Showers}
We analyze the propagation of Askaryan radiation in the lunar regolith. By solving Maxwell's equations in a
dielectric medium such as the lunar regolith one can find the variation of
the radiated signal versus time and location. Similar to \citep{fdtd2}, an illustration of the analysis of the Askaryan radiation in the lunar regolith is shown in Fig.\ref{propag}. Jz and Jr are the vertical and radial components of the current density.  The current is due to the impact of the energetic particles. E(r,z,t) represents the variation of the Electric field in space and time. For Askaryan radiation the equations can be simplified as follows:
In this analysis cylindrical coordinates centred about the cascade are chosen to adjust for longitudinal and radial components. 
The lunar regolith is assumed to be homogenous and the cascade shower
is symmetrical about the shower axis, therefore, the dependency on $\phi$
disappears in Maxwell's equations. \\ \\Furthermore, Askaryan radiation is
highly linearly polarised ( e.g. \citep{saltzberg}).  Therefore its electric fields have components only in the radial direction and along the shower
axis. This would eliminate $E_{\phi}$ and magnetic field components
of $ H_{r}$ and $H_{z}$ so the Maxwell equations reduce to (e.g. \citep{fdtd2}): 
\begin{equation}
\label{eq:3}
\begin{split}
-\partial H_{\phi}/\partial z = \epsilon . \partial E_{r}/\partial t + \sigma . E_{r} \quad \quad \quad \quad  \quad \quad \quad \quad 
\\ \\
1/r . ( \partial(rH_{\phi})/\partial r)=\epsilon . \partial E_{z}/\partial t + \sigma . E_{z} \quad  \quad \quad \quad \quad  \\ \\
 \partial E_{r}/\partial z-\partial E_{z}/\partial r=-\mu  .\partial H_{\phi}/\partial t\quad  \quad \quad \quad \quad \quad \quad 
\end{split}
\end{equation}
Where $\epsilon$ is the dielectric permittivity
and $\mu$ is the magnetic permeability  and $\sigma$ is
the conductivity of the lunar regolith.
The general solution to above equations can be written as separable functions \citep{jackson}:
\begin{equation}
\label{eq:4}
\begin{split}
H_{\phi}(r,z,t)=f_{1}(r) \, . \, f_{2}(z) \, . \, f_{3}(t) \quad  \quad \quad \quad \quad \quad \quad  \quad \quad\\ \\
f_{3}(t)= \exp(j\omega t) \quad  \quad \quad \quad \quad  \quad  \quad \quad \quad \quad \quad  \quad \quad \quad \quad\\ \\
f_{2}(z)=\exp(-\gamma z) \quad  \quad \quad \quad \quad  \quad  \quad \quad \quad \quad \quad  \quad \quad \quad \quad\\ \\
\gamma = \alpha + j\beta \quad \quad  \quad \quad \quad \quad \quad  \quad  \quad \quad \quad \quad \quad  \quad \quad \quad \quad  \quad
\end{split}
\end{equation}
Where $\omega$ is the angular frequency, $\alpha$ is the attenuation constant, $\beta= 2 \pi /\lambda$ is
 the propagation constant and \\
\\
$f_{1}(r)$=$c_{1}$ . $J_{1}(K_{1}r)+c_{2}$ . $Y_{1}(K_{1}r$)\\


$c_{1}$ and $c_{2}$ are constants and  $J_{1}(K_{1}r)$ and $Y_{1}(K_{1}r)$ are Bessel's functions of the first kind and second kind, respectively. \\
Considering the Eq.\ref{eq:3} and the derivatives of the Bessel's functions as:\\
\begin{equation}
\label{eq:derB}
\begin{split}
\partial J_{1}(r)/\partial r = J_{0}(r)- J_{1}(r)/r\\ \\
\partial J_{0}(r)/\partial r = - J_{1}(r) \quad \quad \quad \quad\\ \\
\partial Y_{1}(r)/\partial r = Y_{0}(r)- Y_{1}(r)/r\\ \\
\partial Y_{0}(r)/\partial r = - Y_{1}(r) \quad \quad \quad \quad\\ \\
\end{split}
\end{equation}

The dependence of $E_{z}$ to r is also a sum of the Bessel's functions as:\\ \\
$g(r)=c_{3}.J_{0}(K_{1}r)+c_{4}.Y_{0}(K_{1}r)$\\ 

By solving Eq.\ref{eq:3},  $E_{r} \hspace{2mm} $and$  \hspace{2mm} E_{z}
  \hspace{2mm}$ become: 
\begin{equation}
\label{eq:5}
\begin{split}
E_{r}=\gamma / (j\omega \epsilon + \sigma) \, . \, \exp(-\gamma z) \, . \, \exp(j\omega t) \, . \, f_{1}(r)\\ \\
E_{z}= \hspace{2mm}  \exp(-\gamma z) \, . \, \exp(j \omega t) \, . \, g(r) \quad \quad \quad \quad \quad \quad \\ 
\end{split}
\end{equation}
and $K_{1}$ is obtained as:\\ \\
$K_{1}=(\omega^2 .\mu .\epsilon + \gamma^2 -j \omega .\mu. \sigma )^{0.5}$ \\

The exact solution of $E_{r} \hspace{2mm} $and$  \hspace{2mm} E_{z}$ depends on source excitement, boundary
conditions and electromagnetic properties of the
medium ($\epsilon$, $\mu$ and $\sigma$). The constants in the above equations are identified with these
conditions. In addition, the fields should be physically acceptable so that
fields should have finite values at r,z=0 and at r,z $\to \infty$.  
Equations \ref{eq:5} provide a general picture of the behaviour of the
fields in time and space in a dielectric medium (e.g. lunar regolith) when a cascade 
shower occurs. The electric fields propagate along the shower axis
and are attenuated exponentially. The radial variation of
the electric field is identified by Bessel's function where
decaying field waves develop inside the lunar regolith. The details of
a general solution to Maxwell's equations for this symmetrical geometry can be found in the literature. (e.g. \citep{fdtd1}). 
   
\subsection{Distinction of UHECRv Events in the Lunar Environment}
From the aperture analysis in
section \ref{secap}, it is understood that downward UHECRv events have the most
contribution in detectable cascade showers on the lunar
surface. As discussed in the literature (e.g. \citep{cone} ), Cherenkov cones are created by UHECRv travelling near-parallel to the Moon's surface.  These UHECRv are not affected by total internal reflection so can partially escape the lunar regolith. This is the area where the Askaryan radiation becomes detectable by antenna so it is important that the antenna has horizontal elements parallel to the Moon's surface. If a tripole antenna is used, two crossed dipoles antenna could be positioned in a horizontal plane parallel to the lunar surface. The third vertical element would be also required for the total 3D coverage of space and would be used for DoA estimation and calibration. When the Earth is within the line of sight of the antenna, the radiation from Earth is received in the horizontal plane so it is indeed advantageous that 2 out of 3 dipole elements are positioned in horizontal plane (maximum of the radiation pattern at the zenith) to avoid radio-frequency interference (RFI) as much as possible.\\
For DoA measurement, a 3D coverage of space is required. This can be done using 3 orthogonal antenna elements to detect x, y and z component of arrival signal. If the antenna is positioned on a  moving platform, which is the case for antennas onboard a lunar orbiter, full Stokes parameters need to be calculated to obtain DoA and polarization of arrival signal. (\citep{gonio}). In general, arrival signals have components in 3 dimensions. However, UHECRv radiation is highly linearly polarised and one or two components might be missing so the second set of an antenna in a different position is highly desirable to localise the received signals. Such a basic radio interferometer can provide a degree spatial resolution\citep{rif}.  \\
Various radio signals are detectable in the lunar environment. In table \ref{tab:iv} common radio emissions in the lunar environment are compared with radiation of UHECRv events. Impact of lunar dust, charged particles, and micro-meteorites generate radio pulses  which cover a broad range of spectrum from kHz to GHz regime \citep{GHzimpact}. It has local peaks at kHz regime which are detectable by the antenna, Dust camera, and Langmuir probe\citep{KleinW}. Those events occurring near the antenna generate pulses which could be confused with UHECRv events for the lunar surface observations. Radio emission from planets (e.g. Jupiter and Saturn) peaks at frequencies up to 40 MHz \citep{cassini} and solar bursts have a broad frequency spectrum up to GHz band. These point sky sources can be localised by beamforming. Terrestrial noise includes Auroral Kilometric Radiation (AKR in kHz band) and man-made RFI (kHz-MHz-GHz bands). Their effect is maximised on the Earth's visibility and should be removed by filters and post-data processing. Also, the data should be extracted from the Galactic background noise which is the global dominant noise in MHz band. Considering these types of radio emission in the lunar environment, the importance of triggering the UHECRv events can be understood better. Meanwhile, it should be emphasized that comparing to the Earth-based observation, the Moon enjoys much less man-made RFI and it lacks an ionosphere preventing ionospheric blockage and signal dispersion. Therefore, the lunar surface is a favorite platform to observe UHECRv events. 

\section{Technical Requirements of Future Lunar Radio Experiments}
In this section, we present the basic system requirements of a radio detector for lunar UHECRv events. It is based on the receiver parameters of our analysis which has been compared with those of large radio array such as LOFAR \citep{Scholten} and SKA \citep{ska} in tables \ref{tab:i} and \ref{tab:ii}. While for ground-based observation, the effects of ionosphere could be the main challenge, for space-based observation, the supplying power and data transfer to the Earth will be the main issues. Here we focus on the system requirements of space-based radio detectors. For that, we start with the preliminary design planned for LRX system and briefly present the recent developments of space-based radio astronomy. Furthermore, we discuss how the findings of this article might be applied to the system design of future lunar experiments. 
Fig.\ref{fig:xv} shows the block diagram of a basic space-based radio experiment including the antenna, analog electronics, power supply, data acquisition and data pre-processing unit. Data post-processing will be done in the 
ground-based stations for transmitted data. Power Supply Unit (DC-PSU) provides the required power in terms of DC voltage for the active components of analog and digital units. DC-PSU typically consists of rechargeable batteries which are charged by solar panels. It also includes Power Distribution Unit, DC-PDU, and DC-DC Converters.     
\subsection {Antennas and Analog Electronics}
\begin{figure*} [tbp]
\centering
\includegraphics[width=0.8\textwidth]{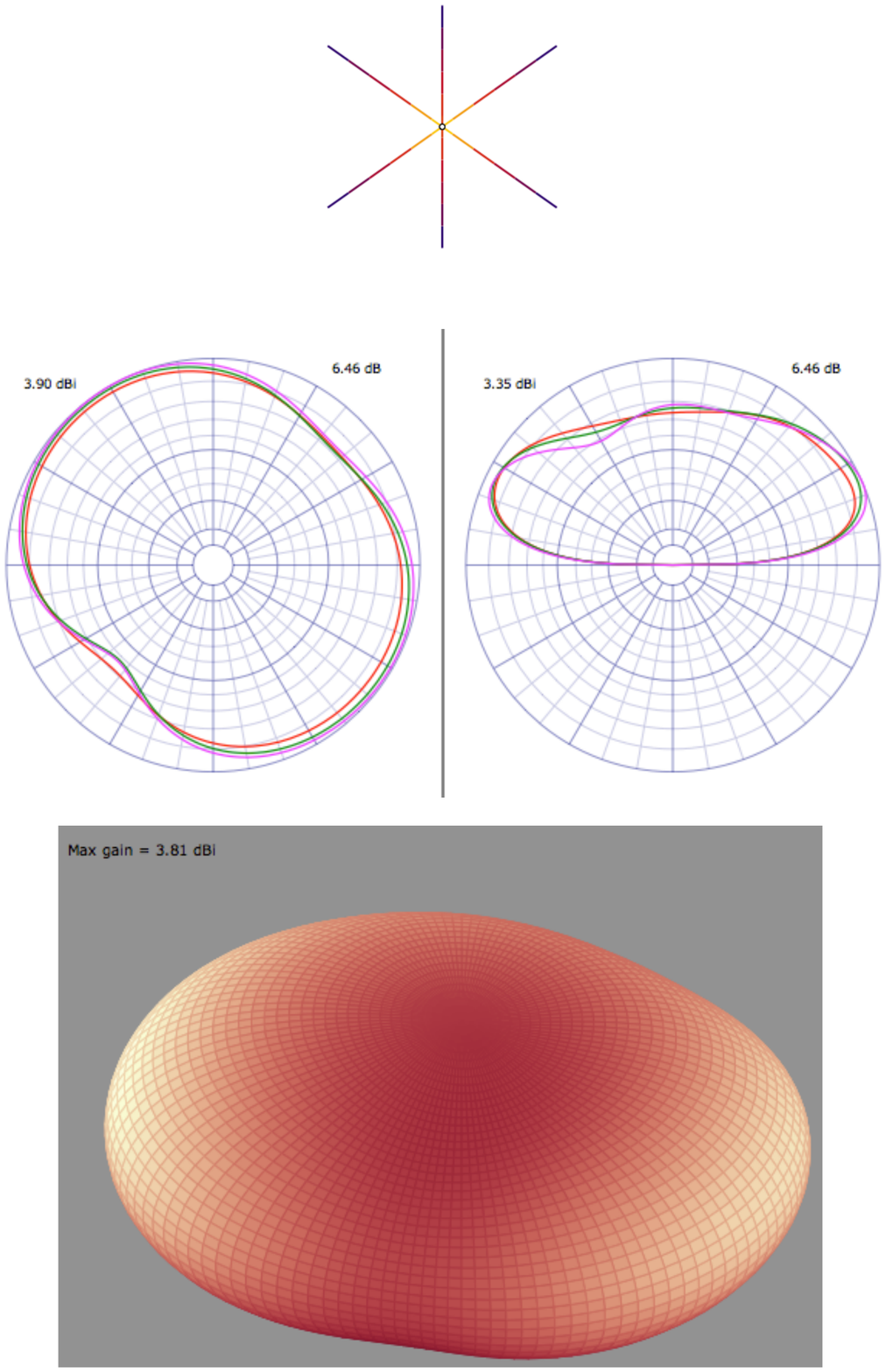} 
\hfill
\caption{\label{FFpattern} Top, the variation of current in tripole arms. The central feed excitations are 1<0 degrees V, \; 1<+45 degrees V and \; 1<-45 degrees V. Bottom, radiation pattern in azimuth (left) and elevation planes. Also the 3D radiation pattern of the tripole in the presence of lunar regolith. The centre of the antenna is 1 m above the Moon's surface and the total length of each dipole is 1 m. Simulation is done using cocoaNEC 2.0 software\citep{cocoanec}}
\end{figure*}

\begin{figure*} [tbp]
\centering 
\includegraphics[width=0.50\textwidth]{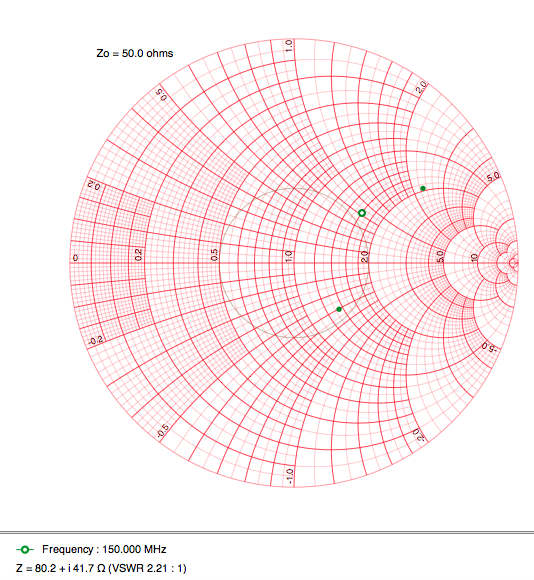}
\hfill
\caption{\label{vswr} An illustration of the Smith Chart for the input impedances and VSWR of the tripole antenna in Fig. \ref{FFpattern}. Green points are the normalized impedances of 140 MHz, 150 MHz and 160 MHz (top). Numbers along the diameter of the biggest circle resemble the normalised resistance. Numbers around the biggest circle resemble the normalised reactance. Number 1 at the centre resembles the VSWR =1 for the perfect antenna impedance matching. The analysis is done using cocoaNEC 2.0 software \citep{cocoanec}}
\end{figure*}

\begin{figure*} [tbp]
\centering 
\includegraphics[width=0.8\textwidth]{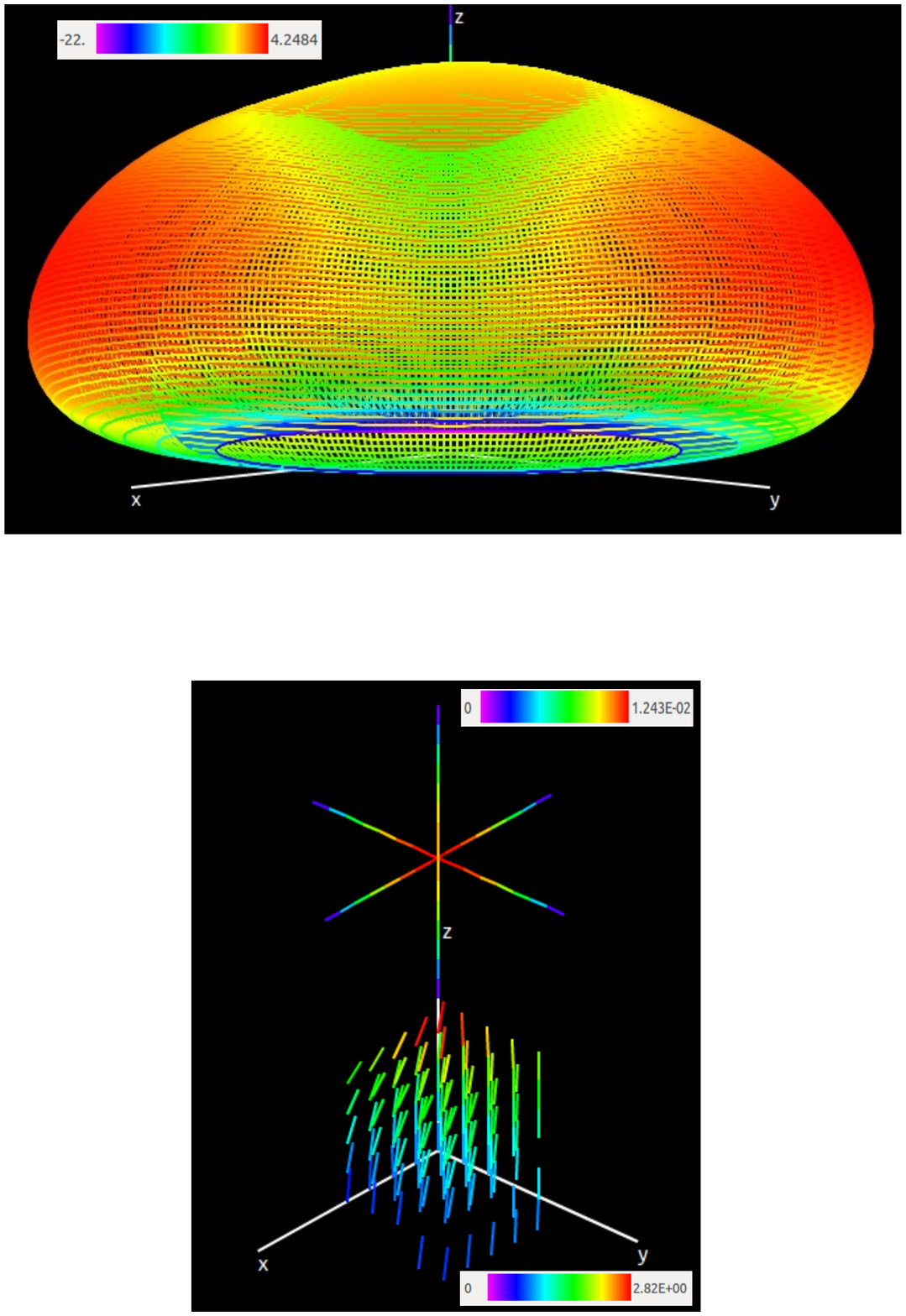} 
\hfill
\caption{\label{fig:NFpattern} Top, simulation of near-field radiation pattern (in dB) of a tripole antenna at 150 MHz in the presence of lunar regolith. Bottom, variation of current in tripole arms. The central feed excitations are 1<0 degrees V, \; 1<+45 degrees V, \; and 1 <-45 degrees V. Also the variation of electric field in the near-field from 0.5 m up to 4 m distance from the tripole. The total length of each dipole is 1 m. Color code on the top is for the antenna current distribution (A) and the one on the bottom shows the variation of the electric field (V/m). Simulation is done using XNEC2c Antenna Software. \citep{xnec} }
\end{figure*}

According to Eq. \ref{eq:1}, the system sensitivity depends on the antenna collecting area and the bandwidth. Antenna collecting area can be increased by including more antennas in the array. It should be noted that an increase of the length of individual antenna elements would affect the performance of antenna and would create side-lobes in the radiation pattern. \\ Although by increasing the bandwidth, the receiver sensitivity is improved but the bandwidth is limited by antenna characteristics. Ideally antenna should have an omnidirectional pattern with the minimum side-lobes over the whole bandwidth to create the maximum aperture on the Moon's surface.  If a resonance antenna is used the standard filters and amplifiers can be used and a matching network is not required. For 150 MHz, resonance (half-wavelength) dipoles with the length of 1m can be used in a tripole structure. For broadband observations at high frequencies, log periodic antennas can be used. For instance, Lunar Orbital Radio Detector (LORD) has two circularly polarized log-periodic spiral antennas for UHECRv observations in a spectrum window of 200-800 MHz. LORD is planned to operate on a polar lunar orbiter spacecraft at 100-150 km altitude for one year and then at 500-700 km for two years \citep{lordd}. \\ Here we present a one antenna candidate for each frequency regime. For 150 MHz, a tripole antenna which consists of three orthogonal dipoles would be a good choice. The length of a resonance dipole at 150 MHz is 1 m which is comparable to the dimension of a lunar lander or lunar orbiter satellites. Therefore the radiation pattern could be affected if the antenna is mounted on a lunar lander or on a satellite. Here a simple analysis of a tripole antenna above the Moon's surface is presented. Simulation is done using cocoaNEC 2.0 software \citep{cocoanec}. The far- field radiation pattern of a tripole antenna in the presence of lunar regolith is illustrated in Fig. \ref{FFpattern}. The total length of each arm is 1m (resonance antenna at 150 MHz) and the 3 dipoles are feed in the centre with for antennas 2 and 3 of the tripole are offset by 45 degrees phases (voltage sources of $1 \angle 0^{o} \;$ V, $1 \angle 45^{o} \;$ V, $1 \angle-45^{o} \;$ V). The centre of the tripole is 1 m above the Moon's surface (half wavelength at 150 MHz). This set up generates a wide beamwidth as shown in Fig. \ref{FFpattern}. The Voltage Standing Wave Ratio (VSWR) of this antenna is illustrated in Fig. \ref{vswr} which shows the best matching at 150 MHz with VSWR of 2.21.  By increasing the length and the distance of antennas from the lunar regolith, side-lobes appear in the radiation pattern. Adding a limited ground plane to the simulated antenna using radial wires of the length of 2m improves the VSWR to 1.71 but the beamwidth reduces. An antenna with limited ground-plane could represent the antenna onboard a lunar lander or lunar satellite orbiter,  the dimesnion and materials of the body of the lander/satellite, however, should be known for exact radiation pattern simulation. The near-field radiation pattern of a tripole antenna in the presence of lunar regolith is illustrated in Fig. \ref{fig:NFpattern}. The antenna has the same set up as for the far-field and simulation is done using XNEC2c Antenna Software \citep{xnec}.  The near-field radiation pattern and the variation of the electric field in the near-field from 0.5 m up to 4 m distance from the tripole is shown in the figure. Also the current distribution across the dipole arms has been shown using colorbars. It can be seen that the near-field radiation pattern is similar to the far-field pattern although the max. gain is slightly higher for near-field patten(4.25 dBi comparing to 3.81 dBi).

For antennas onboard lunar satellites at 1.5 GHz, a helical antenna can be used. Using the formula of a helical antenna in \citep{kraus} the basic parameters of a helical antenna in the presence of a ground plane with dimension of at least 2 wavelengths is calculated. The antenna dimensions (in GHz band), its circular polarization and half-power beamwidth (HPBW), make helical antenna a good choice for an antenna onboard a lunar orbiter satellite for detection of UHECRv events. At 1.5 GHz, a ground plane with dimension of $2 \lambda=40 \;$ cm  and bigger is suitable, which can be fairly approximated with the body of a satellite with this size. For smaller ground planes, the radiation pattern would partially be scattered. The typical helical antenna parameters are summarized in table \ref{helix}. At 1.5 GHz, such an antenna has a circumference of 0.2 m with 5 turns and spacing of 4 cm between turns. The 3dB bandwidth of the antenna is 320 MHz, about 20\% of the central frequency. The antenna operation mode is axial mode and the polarization is circular which can detect both quasi-horizontal and vertical events. The HPBW is 52 degree which  covers the entire lunar aperture for a lunar orbiter experiment at 1000 km distance.  The antenna input impedance of 140 $\Omega$ corresponds to VSWR=2.77 which can be improved by using a matching impedance. For localisation of the events, at least three individual antenna is required. Localisation is part of the event detection and initially excludes other lunar radio emission which is not occurring in the lunar regolith from the likely UHECRv events. 
The analog electronics typically include analog filter and low noise amplifier. If the receiver is exclusively designed for UHECRv detection a band-pass filter is preferred. This will remove the unwanted radiations outside the selected window. For instance, for a 150 MHz observation using a resonance antenna, a bandwidth of 20 MHz is expected. For observations at low frequencies resonance antenna might not be used due to the long length, therefore a matching network would be required. Since matching network limits the system bandwidth, non-matched dipoles have been proposed for broadband observations \citep{nm-dipole}, \citep{Raj}. \\ For broadband receivers, a filter after amplifier is needed to remove the components of harmonics generated in the amplifier. The gain of the amplifier varies based on the gain of the antenna and the distance of antenna from the Moon's surface. Since antenna gain in space-based UHECRv experiments is ideally low, the gain of the amplifier is rather high (more than 20 dB) so the amplifier might be designed in 2-3 stages. This is to meet the receiver sensitivity requirements e.g. the minimum signal level required for the ADCs (analog-to-digital converters). 
 
\subsection {Data Acquisition and Data Processing Unit} 
\begin{figure*}[t]
\centering 
\includegraphics[width=0.7\textwidth]{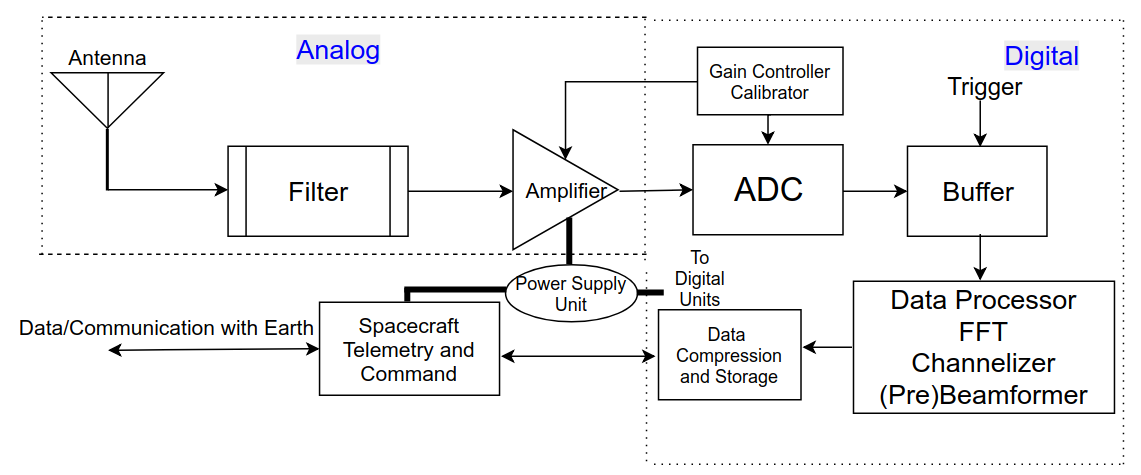} 
\hfill
\caption{\label{fig:xv} A block diagram of basic system requirements for a lunar UHECRv radio experiment}  
\end{figure*}

The digital unit includes the data acquisition (DAQ) and Data Processing Unit (DPU). The amplified signal is digitized by ADCs. The ADC sampling rate should meet the Nyquist limit ($f_s >= 2 f_0$) which $f_s$ is the sampling frequency and $f_0$ is the frequency of observation. 10-bit ADCs are planned for LORD Experiment with $f_s$ of 2 Gsps \citep{lordd}. The dynamic range of the digital receiver can be improved by adding more bits to ADCs at the expense of additional processing time and power consumption. For each run of data taking, the ADC is calibrated and the level of input signal is adjusted by a gain controller. 
In addition to Discrete Fourier Transform (DFT) or cross-correlation between input channels, the digitized data can be channelised in the frequency domain. This, for instance, can be done using a Polyphase Filter Bank(PFB) which has been suggested for a distributed space-based radio array \citep{Raj}. 
Compared to DFT, PFB is more complex and roughly costs 1.5 times more but it produces a relatively flat response across the channels and prevents the leakage of  the signals in the nearby channels \citep{Casper}. This factor is particularly important for RFI Mitigation (RFIM) while detecting weak astronomical signals.  It, however, has been argued \citep{ska} that detection of a UHECRv radio pulse in the order of nanoseconds is problematic after PFB as it splits the radio signals over the multiple frequency bands. In order to prevent that, an invert PFB process with a minor loss of efficiency can be applied to restore the time domain signal for each beam \citep{Singh}. Alternatively, a pre-selection of events can be achieved before PFB in time series data. For instance, digitized time-series data is sent to a temporary memory, the buffer unit (Fig. \ref{fig:xv}). Depending on the duty cycle of the antenna, the main data stream could generate a large volume of data but only those selected by the trigger are stored and sent for processing. 
For a distributed antenna array, data is synchronized using precise onboard clocks. For a single tripole antenna, a common onboard clock for 3 input channels would be needed while for a distributed space-based array,  each satellite needs a clock. The choices of space-qualified clocks for such an array are introduced in \citep{Raj}. It has been also discussed that in the presence of RFI, the Effective Number Of Bits (ENOB) should be greater than 12. For an observing frequency of 150 MHz, this matches to a Signal to Noise Ratio (SNR) around 60 dB and requires a jitter timing of 1 ps and smaller \citep{Raj}. This makes the detectable signal well above the noise level and the likely events can be triggered. Data correlation and Pre-beamforming can be achieved at this stage along with RFI mitigation. Various methods of analog and digital beam-forming and a combination of both analog and digital in a multiport Butler Matrix have been proposed for space communications which also can be used for multi-beam radio telescopes.  For instance, a fully reconfigurable Beam-Forming Network can be done using M $\times$ N control elements of the variable phase shifters and attenuators. The network generates N independent beams from M inputs of antenna elements \citep{spacebeam}.   
In parallel to the channelised data (in the frequency domain), snapshots of time series data can be sampled and those corresponding to the selected events will be stored \citep{KleinW}. The Earth's transfer data rate is usually very limited and varies between tens of kbps to few Mbps(e.g. \citep{Raj}) depending on the Earth visibility and link budget parameters. The volume of stored data can be reduced by compression techniques in the data processing unit before data is sent to the Earth. For lossless compression, the compression ratio of 10 and higher has been reported for space missions \citep{lossless}.

 
\section{Summary and Conclusion}
\begin{figure*}[t]
\centering 
\includegraphics[width=0.7\textwidth]{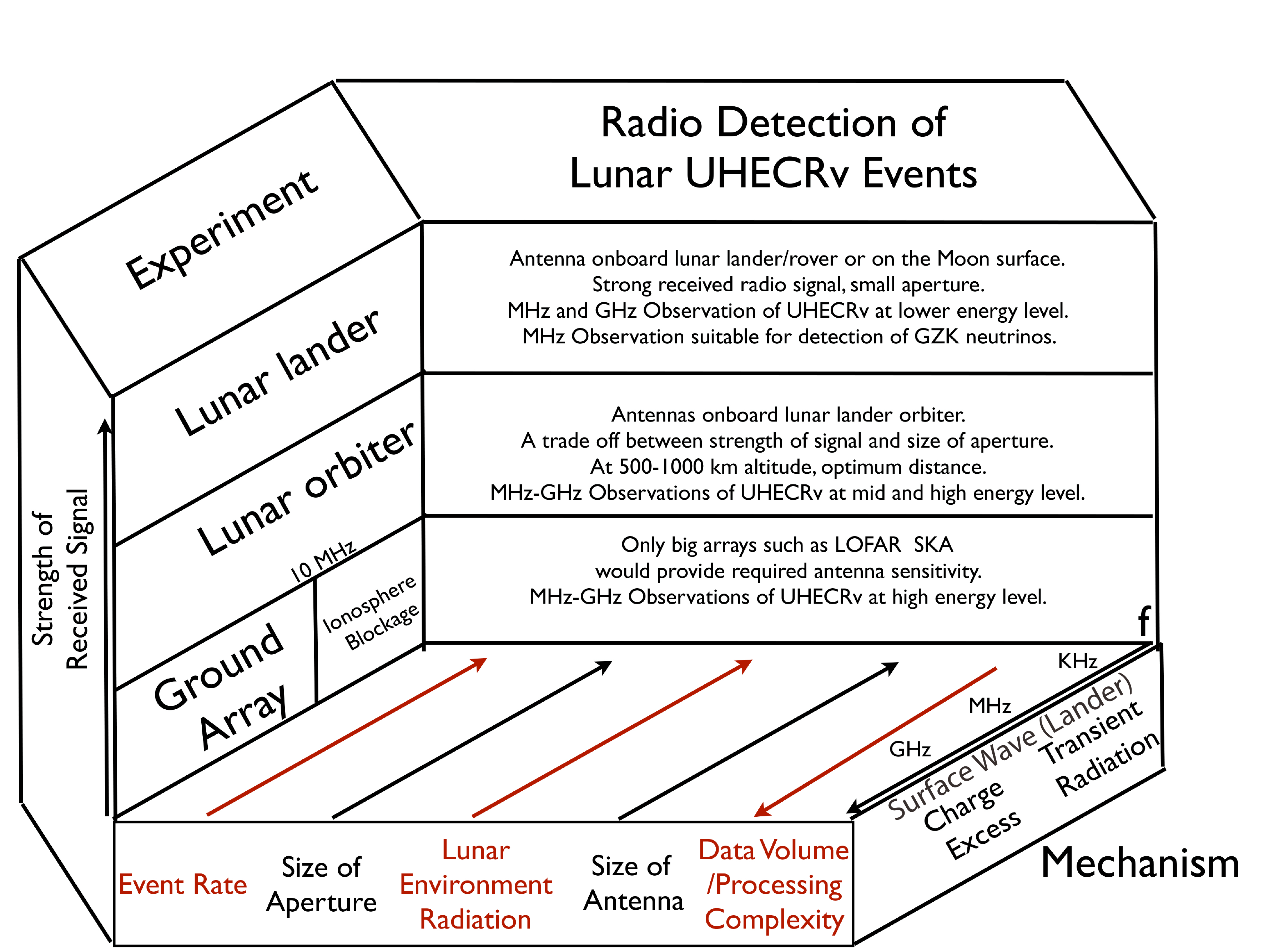} 
\hfill
\caption{\label{fig:xvi} A flowchart of lunar UHECRv radio experiments}  
\end{figure*}

We examine the possibility of a radio detection of lunar UHECRv events in two frequency regimes a 150 MHz and a 1.5 GHz. For the selected frequencies, the analytical method is reliable for the whole range of energy (\citep{ref01}, \citep{ref02}). \\ For future lunar missions, we ran a simulation using a single antenna at different distances from the lunar surface and compared the results with ground-based observations. The method can be applied to an array of antennas on the Moon's surface or multiple antennas onboard lunar orbiters. The results show that the size of UHECRv aperture depends roughly on the square of the wavelength of observation. Therefore the chance of detecting UHERCv events significantly increases in the MHz regime. For the system requirements, the size of the antenna depends on the wavelength of the observation so the antenna will be shorter in GHz regime. However, for the digital receiver, the sampling rate of digitizer depends on the frequency bandwidth which makes the digital processing in the GHz band more complicated. Due to the large volume of digitized data in GHz band, data transfer to the Earth could be particularly problematic for ongoing lunar missions.  Lunar UHECRv events in the low-frequency regime (kHz to few MHz) are also possible and are likely to be influenced by transition radiation as opposed to charge excess mechanism for the MHz and GHz bands. kHz observation, however,  is not practical for lunar missions due to the size of antennas to be deployed in space or on the lunar surface.\\ For selected frequencies, the expected event rates (Fig.\ref{fig:v}, Fig.\ref{fig:viii}) for cosmic rays, GZK neutrinos and TD neutrinos are consistent with the previous numerical and analytical simulations in the literature (\citep{Gusev}, \citep{Stal},\citep{ref01},\citep{ref02}). However, the expected event rate for Z-Burst cosmic neutrinos shows a significant reduction after applying the limits of recent observation (ANITA II, \citep{ANITA2}). This predicts roughly fewer Z-Burst neutrino events by 2 orders of magnitude for both frequencies. For detection of lunar GZK neutrinos, an array of hundreds of antennas on the Moon's surface seems to be the most probable choice, while for higher energy levels (e.g. TD Neutrinos and Z-Burst Neutrinos), lunar orbiter experiments at distances of 500-1000 km would provide the optimum position to observe lunar UHECRv events. \\
We also evaluate the possibility of the detection of neutrinos in the lunar sub-regolith by calculating UHECv apertures. In this paper, we take this into account only for lunar observation at 3 m distance from the lunar surface, and find that an additional contribution of 50\% to 60\% is expected from the sub-regolith. \\ Our preliminary study shows that events occurring in sub-regolith are also detectable as far away as Earth-based observations, since radiation attenuation is very low in the lunar regolith and sub-regolith. This would be an interesting topic for future study to apply electromagnetic properties of the lunar environment in relation to the detection of Askaryan radiation due to the impact of UHECRv. \\ 
We also investigate the propagation of Askaryan radiation in the lunar regolith. Assuming that the shower cascade generates upward and radial components of the current density, the electric fields will be decaying exponentially while propagating towards the Moon's surface. The radial propagation with a pattern of Bessel's functions.  
Fig.\ref{fig:xvi} illustrates a summary of lunar UHECRv radio experiments and the effective parameters. 
\begin{table*}[b]
\centering
\begin{tabular}{l*{6}{c}r}
Radio Emission & Characteristics & Dominant Frequency Spectrum\\
\hline\\
Lunar dust and charged particles& Nearby antenna surface emission& kHz \\
(Micro) Meteorites &Nearby antenna surface emission&MHz-GHz\\
Sky radio sources
(Sun, Planets)& Strong point sources& kHz-MHz\\
Terrestrial noise (AKR,RFI)&On Earth visibility& kHz-MHz\\
Galactic background noise&Global emission&MHz\\
UHECR&Lunar surface(horizontal angles) &MHz-GHz\\
UHECv&Lunar surface and regolith&MHz-GHz\\
\end{tabular}
\caption{\label{tab:iv} Radio emission in the lunar environment}
\end{table*}

\begin{table*}[b]
\centering
Typical parameters of a helical antenna in the presence of a ground plane\\
\begin{tabular}{l|c|c}
\hline\\
Frequency(F) & 1.5 GHz & \\
Bandwidth (BW)&1348 MHz-1668 MHz &320 MHz \\
Number of turns(N)&5\\
Spacing between turns(S)&$0.2\lambda$&4cm\\
Circumference of Helix (C)&$\lambda$&0.2m\\
Input Impedance&$140C/\lambda$&140 $\Omega$\\
HPBW (Degrees)&$ 52 \lambda^{1.5}/[C(NS)^{0.5}]$&52\\
Directivity&$15NC^{2}S/\lambda^{3}$&11.76dBi\\
Operation mode& Axial&Circular polarization\\
\end{tabular}
\caption{\label{helix} Typical parameters of a helical antenna in the presence of a ground plane calculated \citep{kraus} for the central frequency of 1.5 GHz.}
\end{table*}


\appendix
\section{Calculation of the UHECRv Apertures}
\label{A}

\begin{figure*}[t]
\centering 
\includegraphics[width=0.4\textwidth]{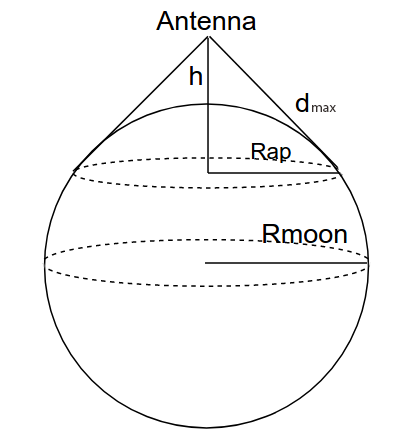}
\hfill
\caption{\label{fig:xvii} An illustration of UHECRv aperture on the Moon's surface. Rmoon: Radius of Moon (1738 km), h: Antenna distance from the Moon's surface, 
Rap: Radius of Aperture, d: Maximum distance of antenna from Aperture}   
\end{figure*}

\begin{figure*}[t] 
\centering 
\includegraphics[width=1\textwidth]{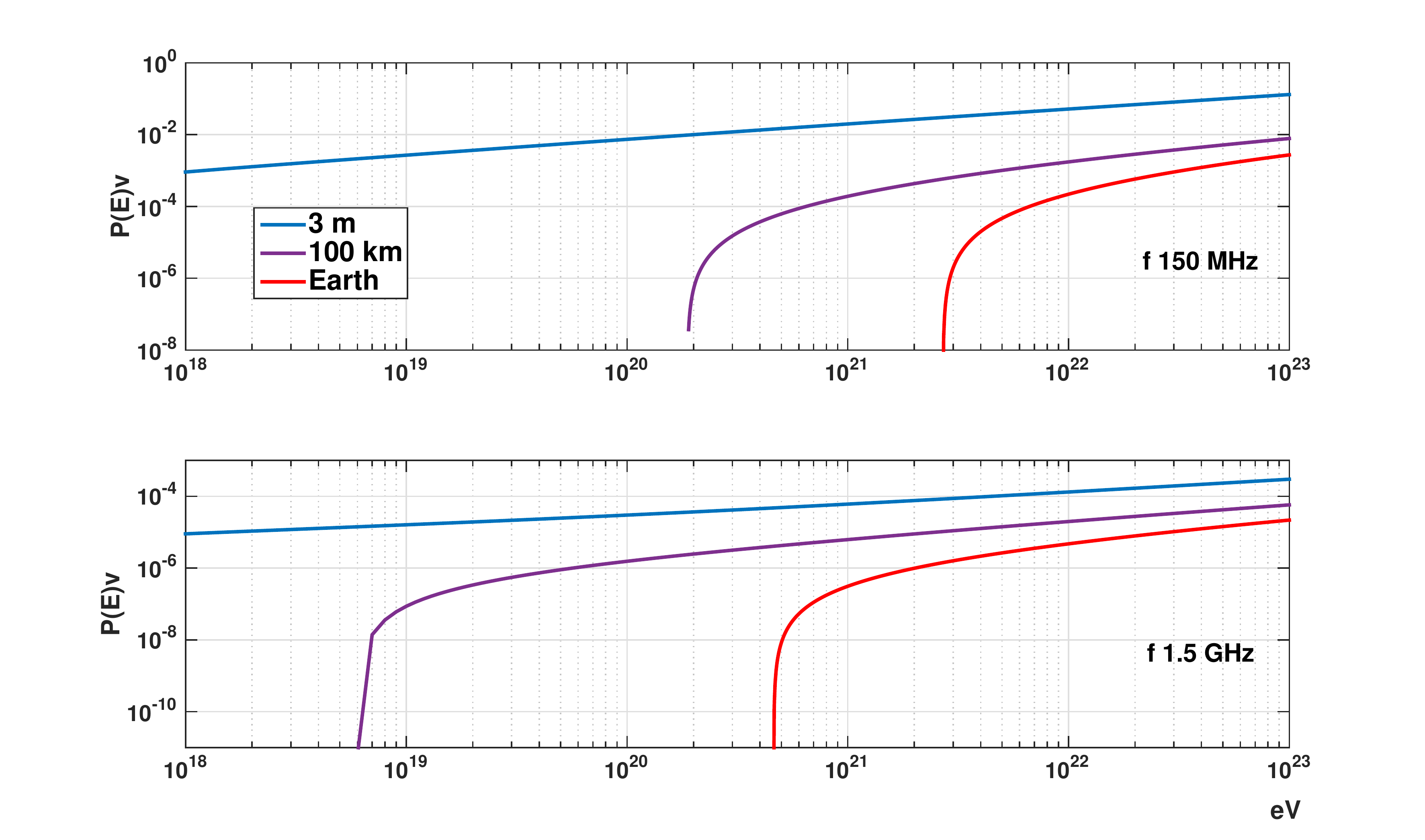} 
\hfill
\caption{\label{fig:xviii} The probability function, P(E), of detection of UHECV events for lunar experiments at frequencies of 150 MHz (top) and 1.5 GHz. P(E)v is plotted for antenna 3m above the Moon's surface (lander), 100 km (lunar orbiter) and Earth-based array. System parameters are shown in table \ref{tab:i} and table  \ref{tab:ii}  }   
\end{figure*}

Here are the modifications made to the analytical methods used in \citep{ref01} and \citep{ref02} for this paper. The main difference is that for Earth-based observations the full Moon's surface is assumed to be illuminated by the antenna, while for lunar orbiters and lunar lander observation only a portion of Moon's surface can be illuminated by the antenna. As opposed to Earth-based observations, the antenna distance from the Moon's surface is comparable to the radius of the illuminated area. Therefore for the illuminated area, the antenna distance is variable. The total aperture ($Ap$) is then calculated by integrating the illuminated area and probability function of the occurrence of UHECRv events over the maximum radius that antenna can illuminate the Moon's surface. The details are illustrated in 
Fig.\ref{fig:xvii} where 
\begin{equation}
\label{eqa1}
\begin{split}
R_{ap} =R_{moon} \, . \, h /(R_{moon}+h) . [(1+2(R_{moon}/h))^{0.5}]\\ \\
Ap(E,f,R_{ap})_{CRv}={\displaystyle \int_{0}^{R_{ap}} A_{0}(r). P(E,f,r) dr} \quad \quad \quad\\ \\
\end{split}
\end{equation}\\  \\
f is frequency and $A_{0}(r)$ is the area on the Moon's surface with radius (r) that is illuminated with antenna radiation beams. $P(E,f,r)$ is a complex function of the probability of the occurrence of UHECv and UHECR events which are adapted from \citep{ref01} and \citep{ref02} respectively. P(E) UHECv for various distances from the lunar surface and at both frequencies of 150 MHz and 1.5 GHz are plotted in 
Fig.\ref{fig:xviii}\\
For calculation of aperture of UHECv events for lunar lander observation, the effect of sub-regolith is also taken into account. \\

$Emax_{Reg.}=0.0845 \, . \,(10^{-6}  (1/d) \, . \, (Es) \, . \,(f_{GHz}) . 10^{-18})  \\ \, . \, (1+((f_{GHz})/2.32))^{-1.23}$ \\
\\
$Emax_{sub}=0.0569 \, . \,(10^{-6}   (1/d) \, . \,  (Es) \, . \, (f_{GHz}) . 10^{-18})  \\ \, . \, (1+((f_{GHz})/2.38))^{-1.23}$\\ \\
Where:
\begin{equation}
\label{eqa2}
\begin{split}
E_{sub}=att \, . \, Emax_{sub} \quad \quad \quad \quad \quad \quad \quad \quad \quad \quad \quad \quad\\
Es=0.2 \times E(eV) \quad \quad \quad for \, Neutrinos  \quad \quad \quad \quad\\
tan\delta= 0.01 \quad \quad \quad  Loss \, \, Tangent \quad \quad \quad \quad \quad \quad \quad\\
\epsilon=3 \quad \quad Dielectric \, \, Constant \, for \, Moon   \quad \quad \quad \quad   \\          
\lambda=3\times 10^8/f  \quad  \lambda \, \, is \,  Wavelength \,   of \,  Observation\\
Ds=\lambda/(\pi \, . \, \delta \, . \, (\epsilon)^{0.5}) \quad  Ds: \,  Skin \,  Depth \quad \quad \quad \quad\\
h_{max}=10 \quad m \quad Thickness \,  \, of \, Lunar \, Regolith \quad \\
att= exp(-h_{max}/Ds) \, Attenuation \,  of \,  Radiation \\ \\
\end{split}
\end{equation} 

Here $Emax_{Reg.} $ is the maximum electric field of Askaryan radiation generated in the lunar regolith. $Emax_{sub}$ is the maximum electric field of Askaryan radiation generated in the sub-regolith. The electric field of sub-regolith events at the antenna point ($E_{sub}$) is calculated with considering the attenuation due to the passage of the signal through 10 m regolith. $ f_{GHz}$ is the frequency of observation in GHz. 
$Emax_{Reg}$ and $Emax_{sub}$ are taken from \citep{fdtd3}. Emax is used in calculation of $P(E,f,r)$. The total aperture then becomes the sum of apertures of regolith and sub-regolith. 

\section*{Acknowledgements}
The work of L.\; Chen is supported by National Natural Science Foundation of China (11573043). \\
We thank the ASR co-editor, Biswajit Paul, and two anonymous reviewers for their constructive comments, which improved the manuscript substantially.  
\clearpage
\section*{References} 
\vspace*{-11.8cm}
\bibliography{asr-bib-rev} 

\begin{thebibliography}{65}
\expandafter\ifx\csname natexlab\endcsname\relax\def\natexlab#1{#1}\fi
\expandafter\ifx\csname url\endcsname\relax
  \def\url#1{\texttt{#1}}\fi
\expandafter\ifx\csname urlprefix\endcsname\relax\def\urlprefix{URL }\fi

\bibitem[{Aab et~al.(2015)Aab, Abreu, Aglietta, et~al.}]{auger-rej}
Aab, A., Abreu, P., Aglietta, M., et~al., 2015. {Improved limit to the diffuse
  flux of ultra-high energy neutrinos from the Pierre Auger Observatory},
  {Phys. Rev. Lett.} Vol.~91. p. 092008.

\bibitem[{Aartsen et~al.(2014)Aartsen, Ackermann, Adams, et~al.}]{icecube2}
Aartsen, M., Ackermann, M., Adams, J., et~al., 2014. {Observation of High-
  Energy Astrophysical Neutrinos in Three Years of IceCube Data },{ IceCube
  Collaboration}, { Phys. Rev. Lett. , 113, 101101}.

\bibitem[{Aartsen et~al.(2013{\natexlab{a}})Aartsen, Abbasi, Abdou,
  et~al.}]{icecube}
Aartsen, M.~G., Abbasi, R., Abdou, Y., et~al., 2013{\natexlab{a}}. {First
  observation of PeV-energy neutrinos with IceCube}, {Phys. Rev. Lett.} Vol.
  111. p. 021103.

\bibitem[{Aartsen et~al.(2013{\natexlab{b}})}]{aj}
Aartsen, M.~G., et~al., 2013{\natexlab{b}}. {IceCube Collaboration, Measurement
  of the high-energy cosmic ray spectrum with IceTop-73}, {Physical Review D88
  }. p. 042004.

\bibitem[{Abraham et~al.(2010)Abraham, Abreu, Aglietta, et~al.}]{CRflux}
Abraham, J., Abreu, P., Aglietta, M., et~al., 2010. {Measurement of the energy
  spectrum of cosmic rays above $10^{18}$ eV using the Pierre Auger
  Observatory}, {The Pierre Auger Collaboration, Phys. Lett. B685}. pp.
  239--246.

\bibitem[{Aminaei et~al.(2013)Aminaei, Klein-Wolt, Chen, et~al.}]{icrc}
Aminaei, A., Klein-Wolt, M., Chen, L., et~al., 2013. {The Prospects of Radio
  Detection of UHECRv on the Moon's Surface}, {in Proceedings of 33rd
  International Cosmic Ray Conference (ICRC), Brasil}. pp. 3358, abstract
  number 0223.

\bibitem[{Aminaei et~al.(2014)Aminaei, Klein-Wolt, Chen, et~al.}]{rif}
Aminaei, A., Klein-Wolt, M., Chen, L., et~al., 2014. {Basic radio
  interferometry for future lunar missions}, {IEEE Aerospace Conference, Big
  Sky, MT, USA.} pp. 1--19.

\bibitem[{Angeletti and Lisi(2013)}]{spacebeam}
Angeletti, P., Lisi, M., 2013. {A Digital Revisitation of Analog Beam-forming
  Techniques for Satellite Multibeam Antennas}, {Conference Paper, 31st AIAA
  International Communication Satellite Systems Conference}. pp. 752--757.

\bibitem[{Arts et~al.(2010)Arts, van~der Wal, and Boonstra}]{nm-dipole}
Arts, M., van~der Wal, E., Boonstra, A.-J., 2010. {Antenna concepts for a
  space-based low-frequency radio telescope}, {ESA Antenna Workshop on Antennas
  for Space Applications, The Netherlands}. Vol.~32. pp. 5--8.

\bibitem[{Askaryan(1965)}]{ref1}
Askaryan, G.~A., 1965. {Excess Negative Charge of an Electron-Photon Shower And
  Its Coherent Radio Emission}, {JETP 14, 441, 1962; also JETP 21, 658.}

\bibitem[{Barwick et~al.(2006)Barwick, Beatty, Besson, et~al.}]{Barwick}
Barwick, S.~W., Beatty, J.~J., Besson, D.~Z., et~al., 2006. {Constraints on
  cosmic neutrino fluxes from the Antarctic Impulsive Transient Antenna
  experiment}, {ANITA collaboration, Phys. Rev. Lett. 96, 171101}.

\bibitem[{Belov et~al.(2015)Belov, Bechtol, Borch, et~al.}]{slac}
Belov, K., Bechtol, K., Borch, K., et~al., 2015. {SLAC T-510: A beam-line
  experiment for radio emission from particle cascades in the presence of a
  magnetic field}, {in Proceedings of 34th International Cosmic Ray Conference
  (ICRC), The Netherlands }. p. 346.

\bibitem[{Bentum et~al.(2009)Bentum, Verhoeven, Boonstra, et~al.}]{olfar}
Bentum, M.~J., Verhoeven, C. J.~M., Boonstra, A.~J., et~al., 2009. {A novel
  astronomical application for formation flying small satellites}, {In 60th
  International Astronautical Congress (IAC), Republic of Korea}. pp. 1--8.

\bibitem[{Bhattacharjee et~al.(1992)Bhattacharjee, Hill, Schramm,
  et~al.}]{Bhatta}
Bhattacharjee, P., Hill, C.~T., Schramm, D.~N., et~al., 1992. {Grand unified
  theories, topological defects, and ultrahigh-energy cosmic rays}, {Phys. Rev.
  Lett. 69, 567}.

\bibitem[{Bray(2016)}]{bray}
Bray, J.~D., 2016. {The sensitivity of past and near-future lunar radio
  experiments to ultra-high-energy cosmic rays and neutrinos}, {Astropart.Phys.
  77}. pp. 1--20.

\bibitem[{Bray et~al.(2014)Bray, Alvarez-Muniz, Buitink, et~al.}]{ska}
Bray, J.~D., Alvarez-Muniz, J., Buitink, S., et~al., 2014. {Lunar detection of
  ultra-high-energy cosmic rays and neutrinos with the Square Kilometre Array
  (SKA)}, {presented in Advancing Astrophysics with the SKA, Italy}. p. 144.

\bibitem[{Carrier et~al.(1991)Carrier, Olhoeft, and Mendell}]{losstangent}
Carrier, W.~D., Olhoeft, G., Mendell, W., 1991. {Physical properties of the
  lunar surface}, {in: G. Heiken et al. (Eds.), Lunar Sourcebook: A User's
  Guide to the Mean. Cambridge Univ. Press, New York}. pp. 475--594.

\bibitem[{CCSDS(2013)}]{lossless}
CCSDS, 2013. {Lossless Data Compression}, {Consultative Committee for Space
  Data Systems (CCSDS) Report Concerning Space Data System Standards}. Vol.~3.

\bibitem[{Cecconi and Zarka(2005)}]{gonio}
Cecconi, B., Zarka, P., 2005. {Direction finding and antenna calibration
  through analytical inversion of radio measurements performed using a system
  of 2 or 3 electric dipole antennas}, {Radio Science, 40, RS3003}.

\bibitem[{Chen(2012)}]{cocoanec}
Chen, K., 2012. {cocoaNEC Reference Manual}, {Open Source Antenna Software,
  Version 2.0}.

\bibitem[{Chen et~al.(2010)Chen, Aminaei, Falcke, et~al.}]{Chen}
Chen, L., Aminaei, A., Falcke, H., et~al., 2010. {Optimized estimation of the
  Direction of Arrival with single tripole antenna}, {Publication of
  Loughborough Antenna and Propagation Conference (LAPC), UK}. pp. 93--96.

\bibitem[{Chennamangalam(2014)}]{Casper}
Chennamangalam, J., 2014. { The Polyphase Filter Bank Technique},{The
  Collaboration for Astronomy Signal Processing and Electronics Research
  (CASPER), Memo 41}.

\bibitem[{Dagkesamanskii and Zheleznykh(1989)}]{DZ}
Dagkesamanskii, R., Zheleznykh, I., 1989. {Radio-astronomy method for detecting
  neutrinos and other elementary particles of superhigh energy}, {Phys. JETP }.
  Vol.~50. p. 259.

\bibitem[{Engel et~al.(2001)Engel, Seckel, and Stanev}]{GZK}
Engel, R., Seckel, D., Stanev, T., 2001. {Neutrinos from propagation of
  ultrahigh energy protons}, {Phys. Rev. D 64, 093010}.

\bibitem[{Gayley et~al.(2009)Gayley, Mutel, and Jaeger}]{ref01}
Gayley, K.~G., Mutel, R.~L., Jaeger, T., 2009. {Analytic Aperture Calculation
  and Scaling Laws for Radio Detection of Lunar-Target UHE Neutrinos}, {ApJ
  706, 1156}.

\bibitem[{Gorham et~al.(1999)Gorham, Liewer, Naudet, et~al.}]{ghzdispersion}
Gorham, P., Liewer, K.~M., Naudet, C.~J., et~al., 1999. {Initial Results from a
  Search for Lunar Radio Emission from Interactions of $10^{19}$ eV Neutrinos
  and Cosmic Rays}, {in Proceedings of the 26th International Cosmic Ray
  Conference, USA}. p. 2.HE.6.3.15.

\bibitem[{Gorham et~al.(2010)Gorham, Allison, Baughman, et~al.}]{ANITA2}
Gorham, P.~W., Allison, P., Baughman, B., et~al., 2010. {Observational
  constraints on the ultrahigh energy cosmic neutrino flux from the second
  flight of the ANITA experiment}, {Phys. Rev. Lett.} Vol.~82. p. 022004.

\bibitem[{Gorham et~al.(2004)Gorham, Hebert, Liewer, et~al.}]{glue}
Gorham, P.~W., Hebert, C.~L., Liewer, K.~M., et~al., 2004. {Experimental Limit
  on the Cosmic Diffuse Ultrahigh Energy Neutrino Flux}, {Phys. Rev. Lett. 93
  (4), 041101 }.

\bibitem[{Grimalsky et~al.(2004)Grimalsky, Berezhnoy, Kotsarenko,
  et~al.}]{GHzimpact}
Grimalsky, V., Berezhnoy, A., Kotsarenko, A., et~al., 2004. {Interpretation of
  the microwave non-thermal radiation of the Moon during impact events},
  {Natural Hazards and Earth System Sciences}. Vol.~4. pp. 793--798.

\bibitem[{Gurnett et~al.(2004)Gurnett, Kurth, Kirchner, et~al.}]{cassini}
Gurnett, D., Kurth, W., Kirchner, D., et~al., 2004. {The Cassini Radio and
  Plasma Wave Investigation}, {Space Science Reviews }. Vol. 114. p. 395--463.

\bibitem[{Gusev et~al.(2006)Gusev, Lomonosov, Pichkhadze, et~al.}]{Gusev}
Gusev, G.~A., Lomonosov, B.~N., Pichkhadze, K.~M., et~al., 2006. {Detection of
  ultrahigh-energy cosmic rays and neutrinos by radio method using artificial
  lunar satellites},{ Cosmic Research}. Vol.~44. pp. 19--38.

\bibitem[{Hu et~al.(2012)Hu, Chen, and Chen}]{fdtd2}
Hu, C.-Y., Chen, C.-C., Chen, P., 2012. {Near-Field Effects of Cherenkov
  Radiation Induced by Ultra High Energy Cosmic Neutrinos}, {Astropart. Phys.}
  Vol.~35. pp. 421--434.

\bibitem[{Horandel et~al.(2009)Horandel, Bahren, Buitink, et~al.}]{lofar}
Horandel, J., Bahren, L., Buitink, S., et~al., 2009. {LOFAR - A new
  experiment to record radio emission from cosmic particles}, {Nucl. Phys. B }.
  Vol. 196. p. 289.

\bibitem[{Jackson(1998)}]{jackson}
Jackson, J.~D., 1998. {Classical electrodynamics}, {Book, John Wiley and Sons,
  Inc., Third edition}.

\bibitem[{Jaeger et~al.(2010)Jaeger, Mutel, and Gayley}]{cone}
Jaeger, T., Mutel, R.~L., Gayley, K., 2010. {Project RESUN, a Radio EVLA Search
  for UHE Neutrinos}, {Astroparticle Physics , Issue 5}. Vol.~34. p. 293--303.

\bibitem[{James and Protheroe(2009)}]{fdtd3}
James, C., Protheroe, R., 2009. {The sensitivity of the next generation of
  lunar Cherenkov observations to UHE neutrinos and cosmic rays},
  {Astropart.Phys.} Vol.~30. pp. 318--332.

\bibitem[{James(2013)}]{jamesrough}
James, C.~W., 2013. {A model for the effects of small-scale surface roughness
  on lunar pulse detection}, {in Proceedings of 33rd International Cosmic Ray
  Conference (ICRC), Brasil }. p. 1052.

\bibitem[{James et~al.(2010)James, Ekers, Alvarez-Muniz, et~al.}]{lunaska}
James, C.~W., Ekers, R.~D., Alvarez-Muniz, J., et~al., 2010. {LUNASKA
  experiments using the Australia Telescope Compact Array to search for
  ultrahigh energy neutrinos and develop technology for the lunar Cherenkov
  technique} , {Phys. Rev. D, 81 (4), 042003 }.

\bibitem[{James et~al.(2011)James, Falcke, Huege, et~al.}]{askaref1}
James, C.~W., Falcke, H., Huege, T., et~al., 2011. {General description of
  electromagnetic radiation processes based on instantaneous charge
  acceleration in endpoints}, { Phys. Rev. E 84, 056602}.

\bibitem[{Jeong et~al.(2012)Jeong, Reno, and Sarcevic}]{ref02}
Jeong, Y.~S., Reno, M.~H., Sarcevic, I., 2012. {Radio Cherenkov signals from
  the Moon: neutrinos and cosmic rays}, { Astrop.Phys., 35(6), 383-395}.

\bibitem[{Klein-Wolt et~al.(2012)Klein-Wolt, Aminaei, Zarka, et~al.}]{KleinW}
Klein-Wolt, M., Aminaei, A., Zarka, P., et~al., 2012. {Radio astronomy with the
  Lunar Lander:opening up the last unexplored frequency regime}, {P\& SS, in
  the special issue SPME}. pp. 167--178.

\bibitem[{Kraus and Marhefka(2002)}]{kraus}
Kraus, J.~D., Marhefka, R.~J., 2002. {Antennas for all Applications}, {3rd
  Edition, USA: McGraw-Hill }.

\bibitem[{Kyriazis(2005)}]{xnec}
Kyriazis, N., 2005. {Xnec2c User Manual}, {Open Source Antenna Software,
  Version 2.3-beta}.

\bibitem[{Marfatia, Mckay and Weiler(2015)}]{marfati}
Marfatia, D., McKay, D., Weiler, T.~J., 2015. {New physics with ultra high
  energy neutrinos }, { Phys. Lett. B., 748}. pp. 113--116.

\bibitem[{Mercurio(2009)}]{Anita1}
Mercurio, B., 2009. {A simulation of the ANITA experiment with a focus on
  secondary interactions}, { Dissertation, The Ohio State University}.

\bibitem[{Nan et~al.(2011)Nan, Li, Jin, et~al.}]{fast}
Nan, R., Li, D., Jin, C., et~al., 2011. {The Five-Hundred-Meter Aperture
  Spherical Radio Telescope (FAST) Project}, {Int. J. Mod. Phys. D 20, 989}.

\bibitem[{Oberoi and Pin\c{c}on(2005)}]{Tformula}
Oberoi, D., Pin\c{c}on, J.-L., 2005. {A new design for a very low
  frequencyspaceborne radio interferometer }, { Radio Science, Vol. 40,
  RS4004}.

\bibitem[{Olhoeft and Strangway(1975)}]{fdtd5}
Olhoeft, G.~R., Strangway, D.~W., 1975. {Dielectric Properties of the First 100
  Meters of the Moon}, {Earth and Planetary Science Letters}. Vol.~24. pp.
  394--404.

\bibitem[{Pugacheva and Shevchenko(2000)}]{moontemperature}
Pugacheva, S.~G., Shevchenko, V.~V., 2000. {The Model of The Moon's Thermal
  Radiation in the Infrared Spectral Ranges (10-12 micron)}, {in 31st Lunar and
  Planetary Science Conference, USA}. p. abstract number 1129.

\bibitem[{Rajan et~al.(2016)Rajan, Boonstra, Bentum, et~al.}]{Raj}
Rajan, R.~T., Boonstra, A.-J., Bentum, M., et~al., 2016. {Space-based aperture
  array for ultra-long wavelength radio astronomy}, {Experimental Astronomy}.
  Vol.~41. pp. 271--306.

\bibitem[{Raybov et~al.(2016)Raybov, Chechin, Gusev, et~al.}]{lordd}
Raybov, V.~A., Chechin, V., Gusev, G.~A., et~al., 2016. {Prospects for
  ultrahigh-energy particle observation based on the lunar orbital LORD space
  experiment}, {Advances in Space Research}. Vol.~58. pp. 464--474.

\bibitem[{Saltzberg et~al.(2001)Saltzberg, Gorham, Walz, et~al.}]{saltzberg}
Saltzberg, D., Gorham, P., Walz, D., et~al., 2001. {Observation of the Askaryan
  Effect: Coherent Microwave Cherenkov Emission from Charge Asymmetry in High
  Energy Particle Cascades}, {Phys.Rev.Lett. 86 , 2802-2805 }.

\bibitem[{Scholten(2007)}]{Scholten}
Scholten, O., 2007. {Optimal Radio Window for the Detection of
  Ultra-High-Energy Cosmic Rays and Neutrinos off the Moon}, {J. Phys.: Conf.
  Ser. 81 012004}.

\bibitem[{Scholten et~al.(2009)Scholten, Buitink, Bacelar, et~al.}]{numoon}
Scholten, O., Buitink, S., Bacelar, J., et~al., 2009. {First results of the
  NuMoon experiment}, {Nucl. Instr. and Meth. V 604, S102}.

\bibitem[{Semikoz and Sigl(2004)}]{TDZ}
Semikoz, D.~V., Sigl, G., 2004. {Ultra-high energy neutrino fluxes: new
  constraints and implications}, {J. Cosmol. and Astropart. Phys. 04. 003}.

\bibitem[{Shoemaker and Morris(1969)}]{regolith10m}
Shoemaker, E.~M., Morris, E.~C., 1969. {Thickness of the regolith.}, {in
  Surveyor: Program results, NASA Special Paper 184. Washington D.C, U.S.
  Government Printing Office}. pp. 96--98.

\bibitem[{Singh et~al.(2012)Singh, Mevius, Scholten, et~al.}]{Singh}
Singh, K., Mevius, M., Scholten, O., et~al., 2012. Optimized trigger for
  ultra-high-energy cosmic-ray and neutrino observations with the low frequency
  radio array. Vol. 664. Elsevier BV, pp. 171--185.

\bibitem[{Sinha and Datta(2012)}]{transitionkhz}
Sinha, K., Datta, P., 2012. {Transition radiation as a tool for radio detection
  of Ultra High Energy Neutrnos}, {in Indian Journal of Radio and Space
  Science}. Vol.~41. pp. 7--16.

\bibitem[{Smirnov and Valovik(2013)}]{fdtd1}
Smirnov, Y.~G., Valovik, D.~V., 2013. {On the Problem of Electromagnetic Waves
  Propagating along a Nonlinear Inhomogeneous Cylindrical Waveguide}, {ISRN
  Mathematical Physics, Article ID 184325, 7 pages, Vol. 2013}.

\bibitem[{St\r{a}l(2007)}]{Stal}
St\r{a}l, O., 2007. {Prospects for Lunar Satellite Detection of Radio Pulses
  from Ultrahigh Energy Neutrinos Interacting with the Moon}, {Phys. Rev.
  Lett., 98(7):071103}.

\bibitem[{ter Veen et~al.(2010)ter Veen, Buitink, Falcke, et~al.}]{askaref2}
ter Veen, S., Buitink, S., Falcke, H., et~al., 2010. {Limit on the
  ultrahigh-energy cosmic-ray flux with the Westerbork synthesis radio
  telescope }, { Phys. Rev. D 82, 103014}.

\bibitem[{Weiler(2003)}]{Weiler}
Weiler, T.~J., 2003. {Physics with Cosmic Neutrinos, PeV to ZeV}, {Int. J. Mod.
  Phys. A}. Vol.~18. p. 4065.

\bibitem[{Williams(2004)}]{Williams}
Williams, D., 2004. {The Askaryan Effect and Detection of Extremely High Energy
  Neutrinos in the Lunar Regolith and Salt}, { Dissertation, University of
  California}.

\bibitem[{Zarka et~al.(2012)Zarka, Bougeret, Briand, et~al.}]{Zarka}
Zarka, P., Bougeret, J.-L., Briand, C., et~al., 2012. {Planetary and
  Exoplanetary Low Frequency Radio Observations from the Moon}, {P\&SS, in the
  special issue SPME, Issue 1}. Vol.~74. pp. 156--166.

\bibitem[{Zas et~al.(1992)Zas, Halzen, and Stanev}]{zas}
Zas, E., Halzen, F., Stanev, T., 1992. {Electromagnetic pulses from high-energy
  showers: Implications for neutrino detection}, {Phys. Rev. D 45, 362 }.

\end{thebibliography}
\end{document}